\documentclass[prx,
 reprint,
superscriptaddress,
 amsmath,amssymb,
 aps,
]{revtex4-2}
\usepackage[]{graphicx}
\usepackage{epstopdf}
\usepackage{bm}
\usepackage{framed}

\usepackage{wasysym} 

\begin{document}
\title{Disorder-aided Early Warning Signals: Predicting Catastrophic Shifts in Athermal Systems}
\author{Tapas Bar}\email{tapas.bar@icn2.cat}
\affiliation{Catalan Institute of Nanoscience and Nanotechnology-ICN2, Campus Universitat Autonoma de Barcelona, Bellaterra 08193, Spain}
\author{Anurag Banerjee}
\affiliation{Universit\'{e} Paris-Saclay, Institut de Physique Th\'eorique,  CEA, CNRS, F-91191 Gif-sur-Yvette, France}
\affiliation{Department of Physics, Ben-Gurion University of the Negev, Beer-Sheva 84105, Israel}
\author{Blai Casals}
\affiliation{Facultat de Física, Universitat de Barcelona, Martí Franquès 1, 08028 Barcelona, Catalonia, Spain}
\author{Gustau Catalan}
\affiliation{Catalan Institute of Nanoscience and Nanotechnology-ICN2, Campus Universitat Autonoma de Barcelona, Bellaterra 08193, Spain}
\author{Javier Rodr\'{i}guez Viejo}\email{javier.rodriguez@icn2.cat}
\affiliation{Catalan Institute of Nanoscience and Nanotechnology-ICN2, Campus Universitat Autonoma de Barcelona, Bellaterra 08193, Spain}
\affiliation{Departament de Fisica, Facultat de Ciencies, Universitat Autonoma de Barcelona, Bellaterra, Barcelona 08193, Spain}
\preprint{To be submitted}

\begin{abstract}
The early prediction of tipping points, distinguished by sudden and catastrophic shifts from stable states, poses a challenging task that would enable us to assess the impending threat across natural and engineered systems. This threat becomes particularly acute in low-fluctuation environments, where tipping occurs through saddle-node bifurcation without prior warning in noise dynamics. In this study, we investigate the tipping point dynamics of avalanche catastrophes in low-fluctuation domain, employing model system like the zero temperature random field Ising model and thermally deposited cobalt films. As the system approaches the tipping point, avalanche activity reveals pronounced critical behaviour, including critical slowing down, variance enhancement, and a growing spatial correlation length--hallmarks that may serve as early warning signals of impending collapse. Crucially, we demonstrate that increasing disorder in the system reduces its vulnerability to catastrophic failure. In highly disorder regimes, these early warning signals emerge well before the transition, thereby providing a large margin for anticipation and mitigation. This key finding suggests a protective role of disorder offering a novel perspective on resilience in complex systems. Our results not only deepen the understanding of tipping phenomena in disorder materials but also have broader implications for forecasting regime shift in diverse real-world systems.
\end{abstract}
\maketitle

Many complex systems such as climate, finances, complex diseases, social and ecosystems start to switch from one state to another state through a cascade process under tiny external perturbations \cite{Book_scheffer2020, Scheffer_Nature09, Scheffer_Science12, Lenton_NatCC11, Lenton_PNAS08, Lenton_PTRSB20}. The inflection points of those chain of events, named tipping points, are extremely hard to predict way before the catastrophe \cite{Book_scheffer2020, grimm2011predicting}. However recent experiments suggest that the `proximity' of a tipping point can be estimated from the recovery response of a perturbed state \cite{Scheffer1_Nature12, Dai_Science12, Carpenter_Science11}. The recovery rates are slower if a system approaches a tipping point, indicating critical slowing down as an antecedent of bifurcation \cite{Lenton_PHRSA2012}. Besides natural complex systems, the prediction of tipping points is gaining attention in real condensed matter systems. For example, strongly interacting Rydberg atom gases manifest early warning signals during the transition from a low-density state to a high-density state under a driving microwave field \cite{Zhang_prl24}. Recently, permutation entropy has been recognized as an anticipation signal of the threshold bifurcation in a complex laser system \cite{Gancio_prl25}. The prediction of the threshold pumping current for stimulated emission in lasers has technological applications for the development of integrated laser systems.

Anticipating critical slowing down from noisy time-series data has become a dominant approach in recent days \cite{Morr_PRX24, Harris_prx24}. The slowing down can equally be detected in the detrended fluctuations of the response function through the increase in variance, decrease in strength of oscillation, the higher correlation of the subsequent states (lag 1), and the increase in auto-correlation time \cite{Scheffer_Science12, Lenton_NatCC11}. Recently, a rapid growth of noise variance and relaxation time has been observed at the charge-density wave transition \cite{Bansal_prl24}. These phenomena occur over an extended temperature range due to the fluctuation-dominated regime, which suggests improved anticipation of the transition.

\begin{figure*}
\centering
\includegraphics[width=1.0\textwidth]{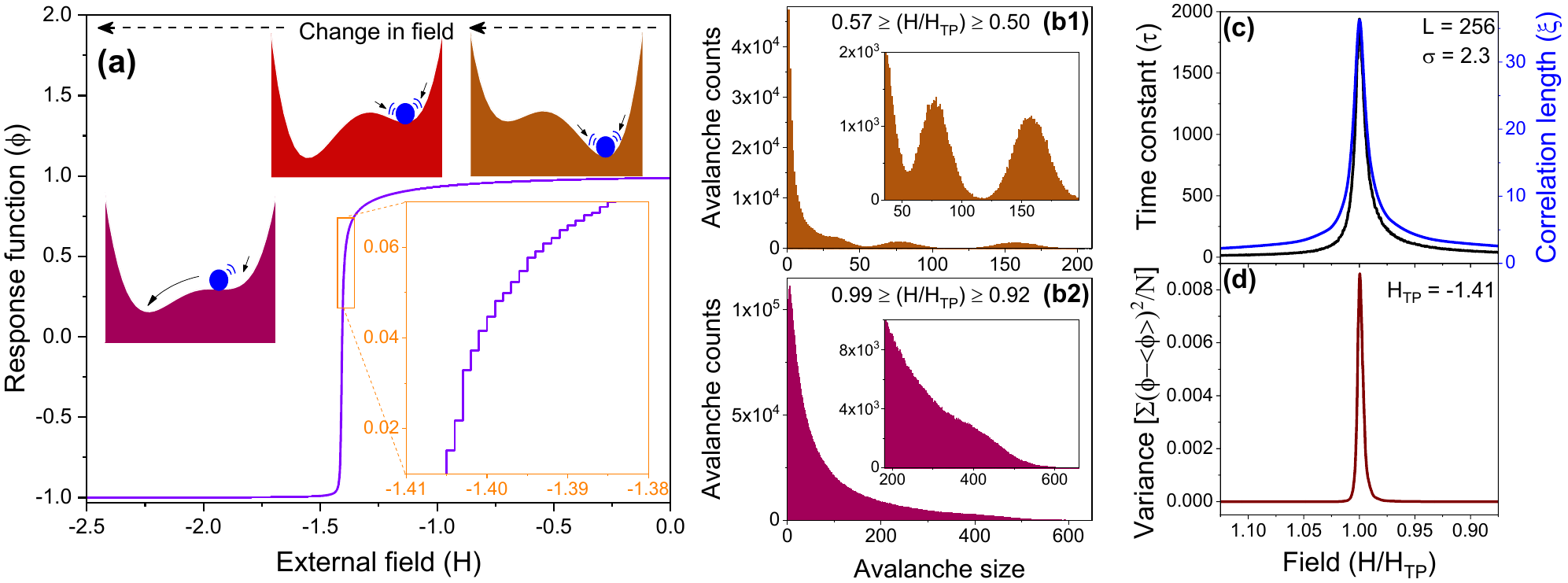}
\caption[0.5\textwidth]{(a) Regime shift carried by successive avalanches due to a change in the external field with a step size of $\delta H = 0.0005$ and corresponding schematic basin of attraction. (b1-b2) Avalanche distribution for small variable intervals ($\Delta H = 0.1$) at different fields before the line of tipping point ($H_{TP} = -1.41$). As the inflection point approaches, large-scale avalanches appearer accompanied by a change in distribution from multimodal to unimodal (insets). The system shows critical slowing down due to the shrinking basin of attraction proximity to the tipping point. The increase in relaxation time constant (c), spatial correlation length (c), and the variance of the response function (d) at the transition point are indications of the slowing down. The data shown here correspond to the decreasing field, in the step of  $\delta H = 0.0005$, applied to $256^3$ system containing disorder strength $\sigma = 2.3$. Here, a regime shift has been presented for up-to-down spin polarization due to an increasing negative field, referred to as the decreasing field. The same phenomenon can be observed for the positive branch of the hysteresis (down to up).}
\label{fig:fig1}
\end{figure*}

However, many systems do not recover in the withdrawal of perturbation; rather, they prefer to stay in the deformed state followed by a hysteretic pathway \cite{Book_Fieguth2016}. For example, avalanches, and glacier melting can not be recovered immediately if we reverse the causing factor \cite{Winkelmann_Nature20, Akesson_NatCom22, Abeouchi_Nature13}. Similar locally irreversible phenomena have also been observed in condensed matter systems in terms of athermal dynamics \cite{Sethna_prl93, Planes_prl01}. The system gets trapped in a metastable local minimum and can only cross the activation barrier under external perturbations, such as a field or temperature change \cite{Bar_prb21}. If there are no external perturbations, the system would never shift to another state even in the vicinity of the tipping point \cite{supplementary}. However, once the system shifts to the deformed state it is impossible to recover because of the large basin of attraction and insufficient fluctuation. Such unidirectional systems are very dangerous- here we can only predict the impending threat and take a measure to minimize the losses associated with the disasters. The prediction is also not antecedent, as the associated fluctuations in a deformed state are uncorrelated (see subsection-C of section-II in the supplementary materials) and therefore unable to show critical slowing down through noise spectroscopy, as observed in the previous mechanism \cite{Prettyman_ERL22, Morr_PRX24}. However, one would still expect to observe critical slowing down in averaged signal and increasing variability of macroscopic response functions due to the basin of attraction at the bifurcation point.

In the context of a conventional Ising ferromagnet below the critical temperature ($T < T_c$), the free energy has two minima, and reversing the external field $H$ switches the ground state from up- to down-spin. In equilibrium, magnetization changes discontinuously at $H = 0$; out of equilibrium, hysteresis arises. As $H$ varies, the double-well potential evolves, and the system can remain trapped in a metastable state if fluctuations are too small to overcome the barrier [Fig. \ref{fig:fig1}(a)]. In the absence of fluctuations, escape occurs only at the inflection point, producing a finite hysteresis width even in quasi-static limits \cite{Book_Nonthermal15}. This ``fluctuation-less'' case can be modeled with the zero-temperature (athermal) dynamics, where clean systems show a single-step catastrophic transition \cite{Sethna_prl93, Planes_prl01}. Real materials, however, contain defects that induce local fluctuations, breaking the cascade into smaller events and producing critical signatures near the transition (see section-I in the supplementary materials). Disordered athermal systems are thus ideal for probing how disorder-driven fluctuations influence critical behavior near the tipping point.

In this article, we examine the resilience of a fluctuation-less system during the transformation from the parent state to the deformed state under a driven external field, both experimentally (cobalt films) and numerically (Ising system). The zero temperature random field Ising model (ZTRFIM) has been considered an athermal model system, primarily applicable for the martensite materials where barrier heights connected to domains flipping (avalanches) are large enough; hence, thermal activation becomes ineffective \cite{Sethna_prl93, Bar_prb21, Bar_prb23}. The fluctuation-less metastable phase can persist close to the unstable singularity and transform unanticipatedly when the system hits the inflection point (tipping point) \cite{Bar_prl18}. At the same time, the magnetic switching in Cobalt (Co) at room temperature is unaffected by thermal fluctuations, making it an ideal example of a magnetic system that behaves athermally \cite{Urbach_prl95_athermal,Salje_prb11,Zapperi_prb98}.

\begin{figure*}
\centering
\includegraphics[width=1.0\textwidth]{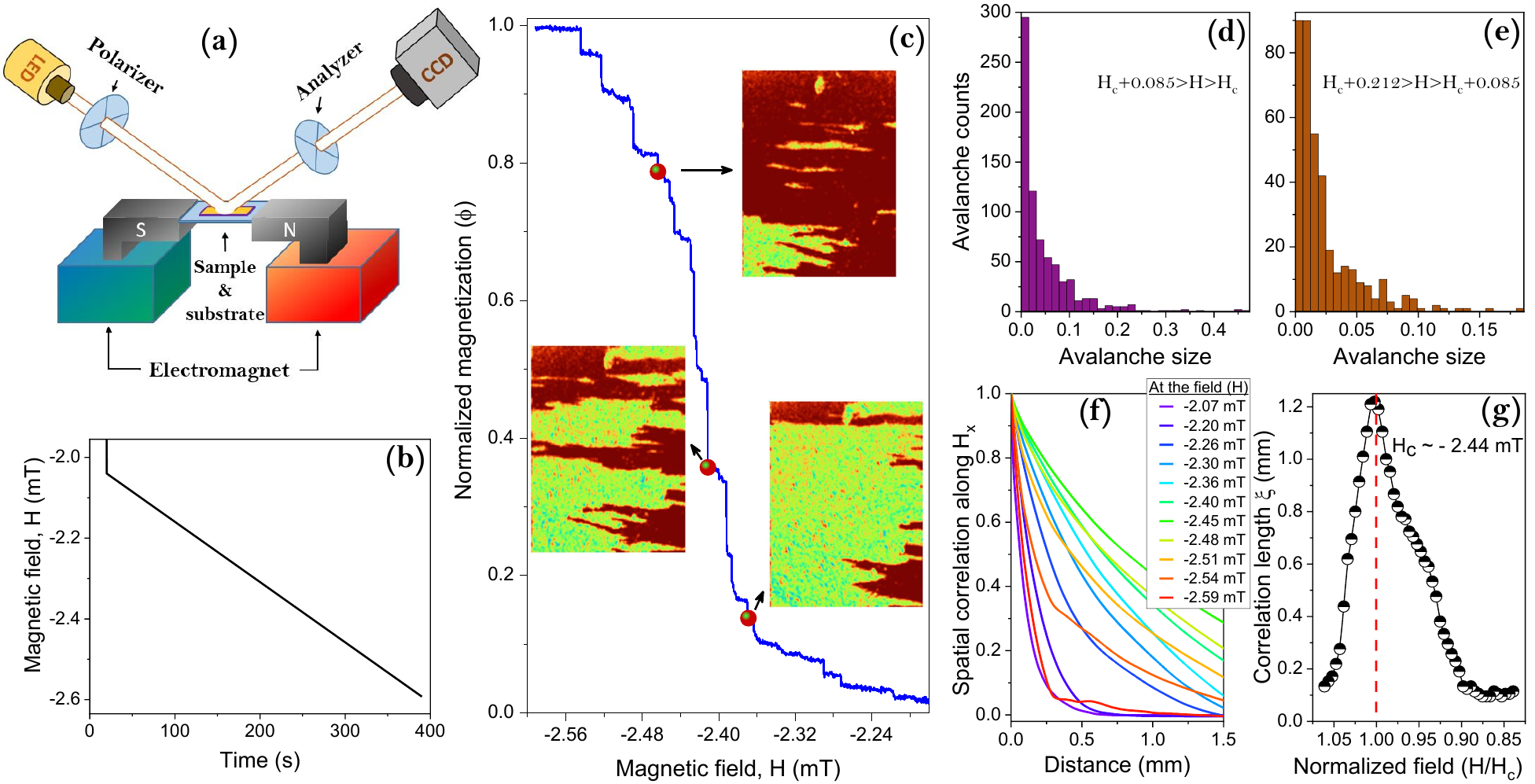}
\caption[0.5\textwidth]{(a) Schematic of the  MOKE microscopy set-up. (b) The time dependent magnetic field sweep at the rate of 1.5 $\mu T/s$ and corresponding magnetic switching through successive avalanches (c). The snapshots of MOKE microscopy measurements have been presented for three different fields. (d, e) Avalanche distribution of magnetizations for 0.085 mT and 0.127 mT field intervals at two different fields before the coercive field ($H_{c} = -2.44$ mT). As field approaches towards transition point, the large size event appears progressively and the distribution of such events becomes unimodal. (f) The spatial correlation of the the MOKE images as a function of distance along the field direction and the characteristic length of the correlation as a function of normalized field, $H/H_{c}$.}
\label{Fig:Expt}
\end{figure*}

In theory, depending upon the strength of spatial inhomogeneities, the avalanche events with changing external field can be classified into three distinct regions \cite{Sethna_prl13, Bar_prb23}. A low-disorder regime where field-induced phase transitions occur between up-spin and down-spin states (or vice versa) through an infinite avalanche, or a spanning avalanche in a finite-size system \cite{Sethna_Nature01}. At a very high disorder regime, phase transition happens through a continual percolation process under the external field change \cite{Sethna_prl13}. Between two extreme disorder regimes, avalanches of small and large sizes co-exist, and the system shows some critical-like phenomena accompanied by an abrupt phase transition \cite{Bar_prb23, Bar_prb21}. At critical disorder $\sigma_c$, the infinite avalanche phase is separated from the critical-like stage through a disorder-induced continuous phase transition \cite{Sethna_Nature01, Sethna_prl93}. In contrast, there is a crossover between avalanche dynamics and percolation dynamics at higher disorder \cite{Bar_prb23}. The percolation-like transition can not be considered a threat in nature. On the other hand, the infinite avalanche appears unanticipatedly and, therefore, is impossible to predict. Fortunately, natural systems never jump infinitely; instead, they show a series of avalanches at the moderate disorder level \cite{Book_scheffer2020, Book_Fieguth2016, Sethna_prl95}. 

Here, we observed critical slowing down, large variance, and an increase in spatial correlation length as early warning signals of tipping points in finite-driven ZTRFIM, which mimic various natural systems. We experimentally verified catastrophic anticipation during magnetic switching in a thermally deposited Cobalt film. Most importantly, we found that the early warning span increases with increasing disorder strength, suggesting that the proximity of the anticipated tipping point becomes more apparent in systems with higher disorder.

\textbf{\textit{Numerical Simulation:}} In the random field Ising model each spin $s_i$ in the lattice points interacts ferromagnetically with their nearest neighbours. The disorder acts as an on-site random field $h_i$ on top of a spatially uniform external field $H$. The Hamiltonian of the model is
\begin{equation}
\mathcal{H} = -J \sum_{i,j} s_is_j + \sum_i[H(t)+h_i]s_i    
\end{equation}
where $J$ (here $J = 1$) is the strength of interaction \cite{Sethna_prl93, Bar_prb21, Bar_prb23}. The simulation has been carried out for a two-state spin configuration ($s_i = \pm 1$) on a 3-dimensional cubic lattice of system size $L^3$ under periodic boundary conditions. The external field $H$ is swept in steps of $\delta H$ from a large positive value (fully up-polarized) to a large negative value (fully down-polarized) and vice versa \cite{Dahmen_prl03, Tadic_prl96, Vives_prl04}. The on-site random field is assigned from a Gaussian distribution $V(h)$, denoted as:
\begin{equation}
V(h) = \frac{1}{\sqrt{2 \pi \sigma^2}} e^{-h^2/(2\sigma^2)}.
\label{eqn:Gaussian}
\end{equation} 

The variance $\sigma$ is the disorder strength. At zero temperature, the spin-flip at each site is regulated by the sign change of the local field. The response function (magnetization) under the perturbation (a tiny sweep of an external field) has been calculated as a function of time and field. At zero temperature, the system transforms through avalanches, where a collection of spins flip upon changing the magnetic field, as shown in Fig. \ref{fig:fig1} (a). 

It is widely recognized that below the critical disorder ($\sigma < \sigma_c$), nearly all the spins flip when the external field reaches the effective coercive field, which is a combination of the external field and the field due to the disorder. Therefore, the avalanche sizes for different disorder configurations are similar, but their positions ($H_c$) vary depending on the random configurations of the disorder. However, a higher disorder strength spatially restricts the correlation function and causes the avalanches to gradually decrease in size \cite{Imry_prb79}. The dynamics of the field-induced transition change from infinite-avalanche (or spanning-avalanche for finite-size system) to critical-like and then to percolative nature with the increase in disorder strength \cite{Zohar_NatSR13, Sethna_Nature01, Sethna_prl93}. 

In the critical-like region, avalanches of different sizes co-exist, and their distribution varies depending on how far away the tipping point is. Figure \ref{fig:fig1} (b1-b2) depicts the size distribution of avalanches within $0.1$ field intervals across various field strengths preceding the tipping point. As the catastrophic threshold approaches, the distribution undergoes a transformation from a short-range multimodal state to a wide-range unimodal state, thereby unveiling a discernible antecedent manifestation of the imminent tipping point. However, those bumps in the avalanche distribution can only be observed for a driven external field with a finite rate and cannot be seen in the adiabatic limit, where the rate is zero, as explained in the supplementary materials \cite{supplementary}. Sometimes, a small bump can be present even in the adiabatic limit at the avalanche size distribution cutoff due to the asymmetric interface height \citep{Laurson_prl24}. 

Nevertheless the precursor indicator observed in the avalanche distribution, the system exhibits a more fundamental phenomenon known as critical slowing down. This is manifested through elevated values of the relaxation time constant, increased spatial correlation, and greater variance of the response function [Fig. \ref{fig:fig1} (c, d)]. Such manifestations during the transition, particularly in the ZTRFIM, have long been investigated within the framework of critical phenomena and are well documented in the literature \citep{Illa_prb11, Handford_JSMTE12, Sethna_prb99}. However, these distinctive characteristics can be interpreted as initial indicators of avalanche tipping points, providing us with the capacity to foresee the proximity of the catastrophic phenomenon \cite{Lenton_NatCC11, Dai_Science12, Carpenter_Science11, Boers_PNAS21, Lenton_PNAS08, Scheffer_Nature09, Scheffer_Science11, Scheffer_Science12, Scheffer1_Nature12, Scheffer2_Nature12}. The prediction has limitations. It applies only to smooth disorder distributions and does not account for abrupt propagation, such as in linear distributions. Moreover, abrupt disorder distributions are not often observed in nature. Detailed calculations of the variance, spatial correlation, relaxation time constant, and the limitations of the model under varying disorder distributions are presented in the supplementary materials \cite{supplementary}.

\begin{figure*}
\centering
\includegraphics[width=0.7\textwidth]{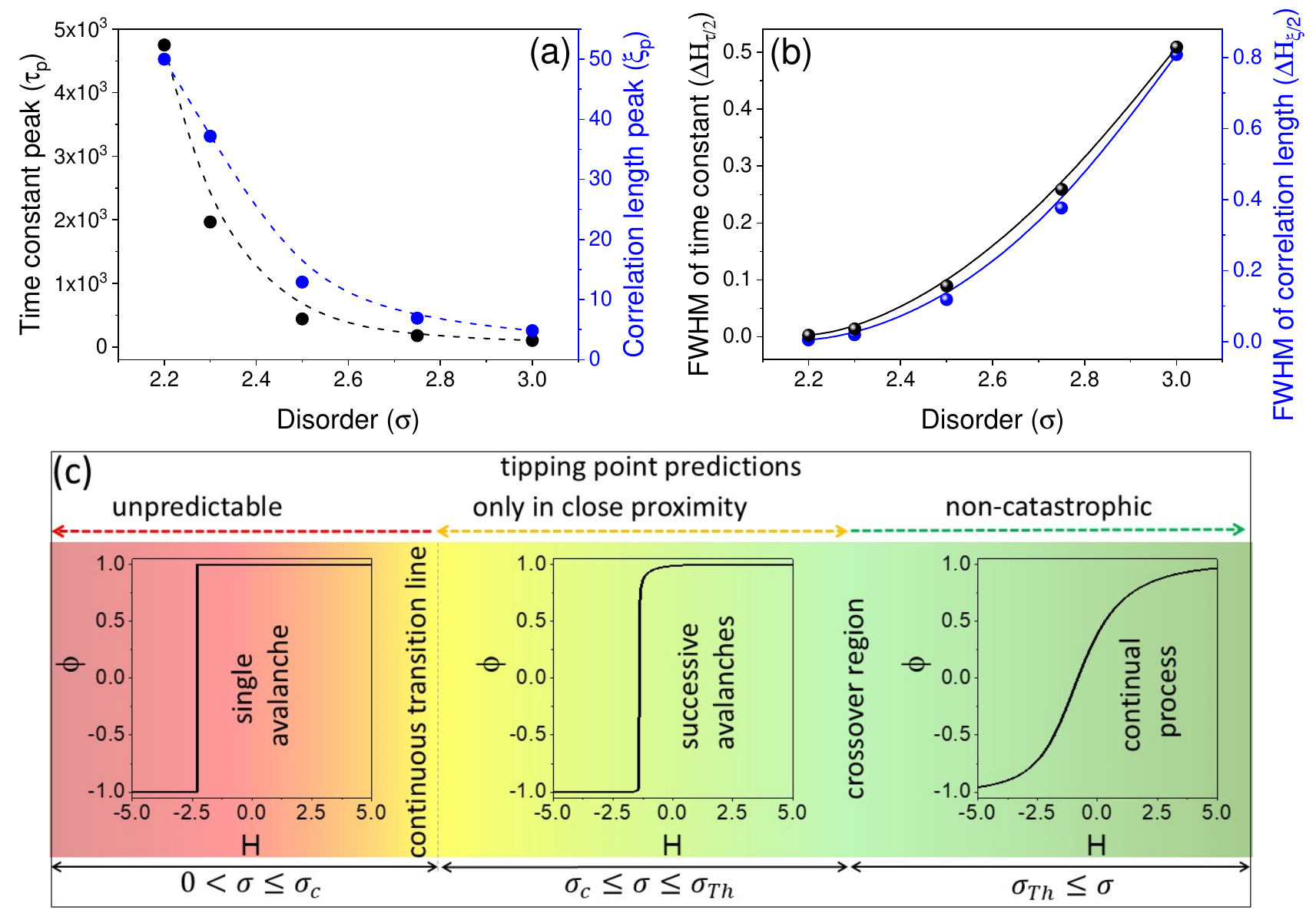}
\caption[0.5\textwidth]{(a) The maximum value ($\tau_p$) and (b) the full width at half maximum ($FWHM$) of the correlation length and the time constant peaks as a function of disorder strengths. (c) Tipping point prediction possibility diagram on the disorder line of ZTRFIM.}
\label{fig:fig3}
\end{figure*}

\textbf{\textit{Experiments:}} To verify the early signals of catastrophic failure, we performed tabletop experiments on the avalanche phenomenon in an elemental magnetic metal (Cobalt). We used the Magneto-Optical Kerr Effect (MOKE) microscopy to investigate the magnetization dynamics under a small linear change in the magnetic field [Fig. \ref{Fig:Expt}(a)] \cite{Qiu_RSI00, Puppin_PRL00}. The spin polarization was detected by measuring the change in polarization of the reflected light from a magnetic sample placed in a magnetic field. This technique allows us to directly measure the Barkhausen avalanches of ferromagnetic films \cite{Puppin_PRL00,Kim_PRL03,Ryu_NatPhys07,Kim_NatCom20}. 

In the experiment, a 100 nm Co film was thermally deposited on a glass substrate. The sample was mounted on a rotational stage situated at the center of the electromagnetic poles. A polarized light beam illuminated the sample at an incident angle of approximately \(45^\circ\). The reflected light was imaged by a CCD camera after passing through an analyzer \cite{Qiu_RSI00, Puppin_PRL00, Kim_PRL03}. The MOKE microscopy measurements were restricted to the longitudinal geometry, where the in-plane magnetic field is parallel to the magnetization vector.

Here, we varied the magnetic field linearly by adjusting the electrical current of the electromagnet and simultaneously captured the MOKE images as a function of the magnetic field during the ferromagnetic switching of the sample [Fig. \ref{Fig:Expt}(b, c)]. The transformation fraction of the image represents the normalized magnetization of the sample. The magnetic switching of the entire Co film evolves through a series of distinct magnetic domain flips, indicating an avalanche-like transformation [Fig. \ref{Fig:Expt}(c)].

To increase the resolution of our data processing, we divided the total image into four segments to ensure that small avalanches were not masked by averaging. To obtain a good statistical distribution, the measurements were repeated multiple times (40 cycles) to achieve a large number of jumps. The avalanche size distribution of normalized magnetizations for two small field intervals prior to the coercive field (\(H_c = -2.44\ mT\)) is depicted in Figure \ref{Fig:Expt}(d, e). As the external field come closer to the coercive field, the range of avalanche sizes becomes extensive, as observed in the numerical simulation of the RFIM [Fig. \ref{fig:fig1} (b1-b2)]. The modes of the distribution in Figure \ref{Fig:Expt}(e) are not as clear as in the simulation [Fig. \ref{fig:fig1} (b1)]. 

There have been numerous studies on avalanche size distribution integrated over all magnetic fields, aiming to address a fundamental aspect of condensed matter physics—the nature of self-organized criticality \cite{Sethna_Nature01,Sethna_prl93,Urbach_prl95,Zapperi_prl97,Durin_prl00,Puppin_PRL00,Kim_PRL03,Ryu_NatPhys07}. Proper scaling requires averaging the total number of events over the entire hysteresis loop (from \(H = +\infty\) to \(-\infty\) or vice versa) \cite{Sethna_prl95, Book_Sethna06}. Several studies have also been conducted for small field windows along the hysteresis loop, which show a change in the scaling exponent as the field approaches coercivity \cite{Im_apl09}, and this change might be considered an indicator of the tipping point. However, the large avalanches that consistently occur near the transition are definitely a signature of the tipping point [Fig. \ref{Fig:Expt}(d, e)] \cite{Urbach_prl95,Zapperi_prl97}. Needless to say, any model, including the random field Ising system, is unlikely to be effective at predicting avalanches one by one \cite{Book_Sethna06}.

Therefore, in this article, we rather focus on macroscopic indicators, such as spatial correlation and variance, which are independent of microscopic events and have greater predictive capacity. Figure \ref{Fig:Expt}(f) shows that the spatial correlation of the magnetization vectors decays with the distance between them, and this decay becomes slower around the transition. Consequently, the correlation length increases at the transition [Fig. \ref{Fig:Expt}(g)]. Along with these indications, an increase in variance around the transformation has also been observed and reported in the supplementary materials \citep{supplementary}. The detailed calculations of variance and correlation length are explained in the supplementary materials \citep{supplementary}. 

\begin{figure*}
\centering
\includegraphics[width=1.0\textwidth]{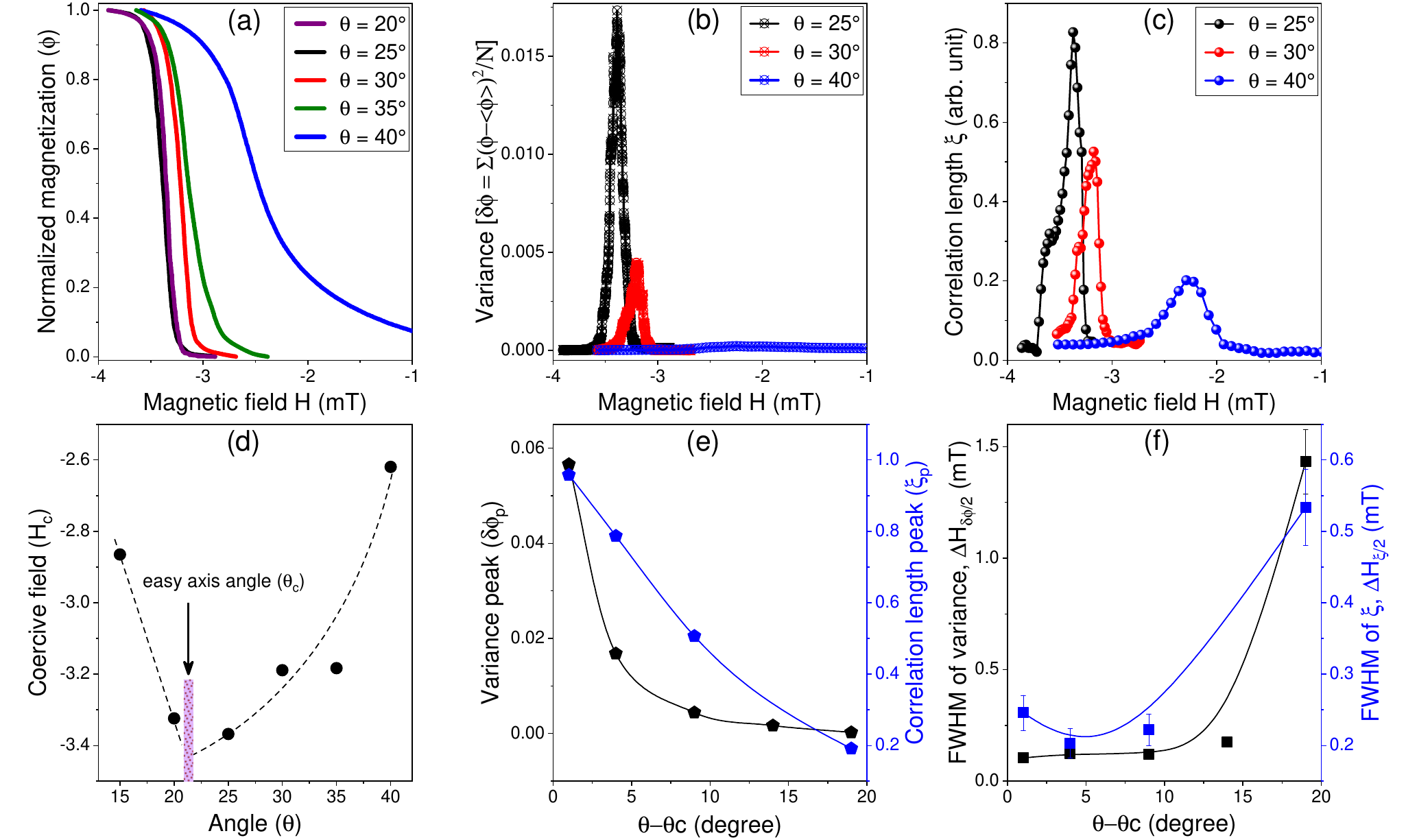}
\caption[0.5\textwidth]{(a) Normalized magnetization as a function of the external magnetic field applied at different angles $\theta$ with respect to the sample axis. Variances were estimated from multiple measurements for each angle, with some examples shown in (b). (c) The corresponding calculated correlation lengths for those angles. (d) The coercive field as a function of the measurement angle. The shaded region indicates the easy-axis angle ($\theta_c \approx 21^\circ$) of magnetization, including the possible error margin. (e) The peak values of variance and correlation length as functions of the relative angle to the easy-axis angle. (f) The increasing broadness of the early warning signals, expressed in terms of the full width at half maximum (FWHM) of variance and correlation length, as a function of $\theta - \theta_c$.}
\label{fig:fig4}
\end{figure*}

Briefly, the critical signatures that start to showcase before the catastrophic failure in the generalized fluctuation-less (zero temperature) Ising model also qualitatively holds for the magnetic avalanches in a elementary magnetic material (Co) way below the Curie temperature \cite{Sethna_Nature01,Kim_PRL03, Ryu_NatPhys07,Lemerle_prl98,Kim_NatCom20,Urbach_prl95_athermal}. Therefore, we can argue that such signs can be considered as early warning indicators of impending catastrophic failure in systems with low fluctuations \cite{Lenton_NatCC11, Dai_Science12, Carpenter_Science11, Boers_PNAS21, Lenton_PNAS08, Scheffer_Nature09, Scheffer_Science11, Scheffer_Science12, Scheffer1_Nature12, Scheffer2_Nature12}.

\textbf{\textit{Effect of disorder:}} The anticipated proximity to catastrophic failure varies in accordance with the intensity of the disorder. To explore this, we have numerically calculated the time constant and spin-spin correlation over the full range of fields for different disordered systems. Detailed simulations and calculations are provided in the supplementary material \citep{supplementary}. In the domain of field-driven transitions within the successive avalanche region, both the relaxation time constant and spin correlation increase for all disorder values. However, it is noteworthy that while the peak height of both time and length scales decreases, the peak width broadens with increasing disorder for both cases [Fig. \ref{fig:fig3}(a, b)], as expected. Smaller peak values indicate a lesser threat to the tipping point, whereas a broader peak begins to manifest a warning signal well before the onset of a cascade catastrophe. Therefore, forecasting becomes more feasible for systems with higher disorder levels. This feasibility surges until a particular disorder threshold, where the two branches of the saddle-node bifurcation converge, rendering the transition non-hysteretic. In this scenario, by definition, the transition point no longer qualifies as a tipping point for a noise-free system \cite{Ashwin_PHRSA2012, Lenton_NatCC11}. Conversely, within the signal avalanche region (\(\sigma < \sigma_c\)), the catastrophic occurrence cannot be attributed to a cascade effect; instead, it seems to arise unexpectedly at the tipping point. Fortunately, the majority of intricate natural systems do not exhibit such phenomena \cite{Book_scheffer2020}. Figure \ref{fig:fig3}(c) showcases the potential for predicting tipping points across the entire disorder spectrum within the field-induced transition in ZTRFIM.

In experiments, precisely controlling and characterizing the strength of disorder across multiple samples is challenging. However, we can qualitatively study the disorder dependence of the samples by performing MOKE measurements on another comparatively thin Cobalt sample (grown under identical conditions) with the magnetic field applied at an angle  ($\theta$) to the sample axis. We observe that, depending on the field angle relative to the axis of magnetization, the switching can be abrupt or smooth, a behavior consistent with increasing disorder. When an external field is applied at an angle $\theta - \theta_c$, where $\theta_c$ is the easy axis angle of the anisotropic ferromagnet, the magnetostatic Zeeman energy smears the abrupt transition in a manner analogous to how disorder restricts the correlation length. Therefore, one can phenomenologically map the disorder strength $\sigma$ to the angle of the field direction relative to the easy axis, $\theta - \theta_c$.

Here, Fig. \ref{fig:fig4}(a) shows that, as the angle increases from the easy-axis angle $\theta_c$, the transitions become broader. A similar change in the shape of the magnetization curve during switching has been observed in the random-field Ising system when the disorder strength is increased \citep{Bar_prb23}. Most importantly, the variance and correlation length peaks become wider, and the peak height diminishes as the magnetic field direction deviates further from the easy axis of magnetization [Fig. \ref{fig:fig4}(b-c)]. The easy axis angle ($\theta_c \approx 21^\circ$) has been estimated from the coercive field versus angle of measurement graph [Fig. \ref{fig:fig4}(d)]. The intensity and proximity of the switching threat are evaluated from the peak value and the broadness of the variance and correlation length peaks. Figure \ref{fig:fig4}(e) shows that the transformation becomes continuum at higher relative angles $\theta - \theta_c$. At the same time, warning signals can be observed far earlier than the actual switching event [Fig. \ref{fig:fig4}(f)].

Therefore, Figures \ref{fig:fig3} and \ref{fig:fig4} effectively demonstrate that with increasing disorder, the capacity to identify the threat earlier becomes more pronounced. However, we acknowledge that the ZTRFIM model presented in this work does not explicitly incorporate the angle-dependent magnetic field or the anisotropic properties of the material. Experimental data suggests that a magnetic field applied away from the easy axis can induce a transverse component of the field with a strong interplay with random fields. Developing a detailed model and conducting an in-depth analysis of such a scenario remain important challenges for future research.

\textbf{Conclusion:} We confirm that critical-like signatures, such as increases in correlation length, critical slowing down, and variance enhancement, offer a powerful framework for predicting impending threats and measuring the breaking point due to external causes. The complex interplay of activation barriers, disorder, and fluctuations provides these critical indicators, collectively contributing to our understanding of the critical phenomena at the tipping points within the framework of abrupt phase transition in various materials \cite{Zapperi_prl97,Bar_prl18,Bar_prl20,Book_Levin23}. The hallmarks of critical behavior in abrupt transitions can be associated with well-established spinodal instability, offering mechanisms applicable for predicting tipping points across diverse complex natural systems  \cite{Bar_prl20, BookChap_Klein01, Book_Levin23}. Additionally, the thermodynamic quantities that are commonly measured in conventional statistical physics for such instability may be used to predict the threats \cite{Yen_ComPhys23}.    

The study further sheds light on the diverse trajectories that systems undertake during transformations, where the nature of transition varies from cascading to abrupt, influenced by the presence of disorder. The interplay between the correlation length and disorder-induced local fluctuations provides insights that resonate with a wide range of disordered systems in nature. From the predictive indicators, our results suggest that a reduction of disorder makes the system more critical and demands for urgent measures to tackle the threat. First and foremost, this article highlights that the threat can be mitigated by externally introducing defects into the system. This knowledge not only enhances our ability to anticipate and manage potential crises in condensed matter systems but also provides insight into the behavior of various complex natural systems \cite{Book_Levin23}.

\section*{Acknowledgments}

It is a pleasure to thank Alvaro Corral and Jordi Bar\'o i Urbea for critical comments , suggestions and discussion. We acknowledge Grant TED2021-131363B/I00 funded by MICIU/AEI/ 10.13039/501100011033 and by the ``European Union NextGenerationEU/PRTR". T.B. acknowledges Juan de la Cierva post-doctoral fellowship. The ICN2 is funded by the CERCA program/Generalitat de Catalunya. The ICN2 is supported by the Severo Ochoa program of MINECO (Grant SEV-2017-0706). TB J.R.-V acknowledge support from 2021SGR-00644 funded by AGAUR. A.B. acknowledge support from the Kreitman School of Advanced Graduate Studies and European Research Council (ERC) grant agreement No. 951541, ARO (W911NF-20-1-0013). The computations were performed on the BGU cluster.

\newpage

\begin{center}
\underline{\textbf{\large Supplemental Material}}
\end{center}

In the first section of the Supplemental Material, we present the motivation behind the chosen model and outline the question of interest in simple terms. In second section, we discuss the numerical simulation procedure and corresponding analysis, focusing on phase ordering dynamics with and without external temporal field fluctuations. A detailed description of the spatial spin-spin correlation in the random field Ising model is presented. In the final section, we analyze the experimental data and describe the methods used to extract spatial correlation and variance. 

\section{Tipping point and athermal dynamics}

In a conventional Ising ferromagnet, where below the critical temperature $T < T_c$ the free energy has two minima, changing the external magnetic field $H$ from positive to negative switches the ground state from up-spin to down-spin configurations. In equilibrium, this produces a discontinuous change in magnetization at $H = 0$; in non-equilibrium, hysteresis appears. Under a change in the external field, the double-well potential changes as shown in Fig.~1(a) (main text). The system prefers the global minimum, but a potential barrier between the two minima prevents it from escaping, leaving it stuck in a quasi-equilibrium, metastable state due to ineffective fluctuations near the local minimum \cite{Book_Nonthermal15}. The system can escape the metastable minimum only when the fluctuations are comparable to the barrier height. In a fluctuationless system, escape is impossible unless the inflection point of the local free-energy minimum is reached---at which point the double-well potential collapses into a single well, and the system exhibits a finite (maximum) hysteresis width even in quasi-static dynamics. Such a transition can be observed in the dynamics of phase transitions using the fluctuationless Ginzburg--Landau--Langevin equation for a non-conserved order parameter \cite{Bar_prl18}.

Microscopically, fluctuationless phenomena have long been studied using zero-temperature (athermal) dynamics, where a disorder-free, clean system undergoes a sudden, single-step catastrophic transition \cite{Sethna_prl93}. In real systems with finite defects, the catastrophic shift is no longer a single-step function. Local fluctuations due to disorder restrict cascade effects and exhibit critical signatures just before the transition. Therefore, disordered athermal systems are ideal for studying the effect of local disorder-induced fluctuations on the critical signatures at the inflection point of the double-well potential. In this context, the zero-temperature random-field Ising model (ZTRFIM) has been used as a model athermal system to examine proximity to the tipping point.

\begin{figure*}
\centering
\includegraphics[scale=0.45]{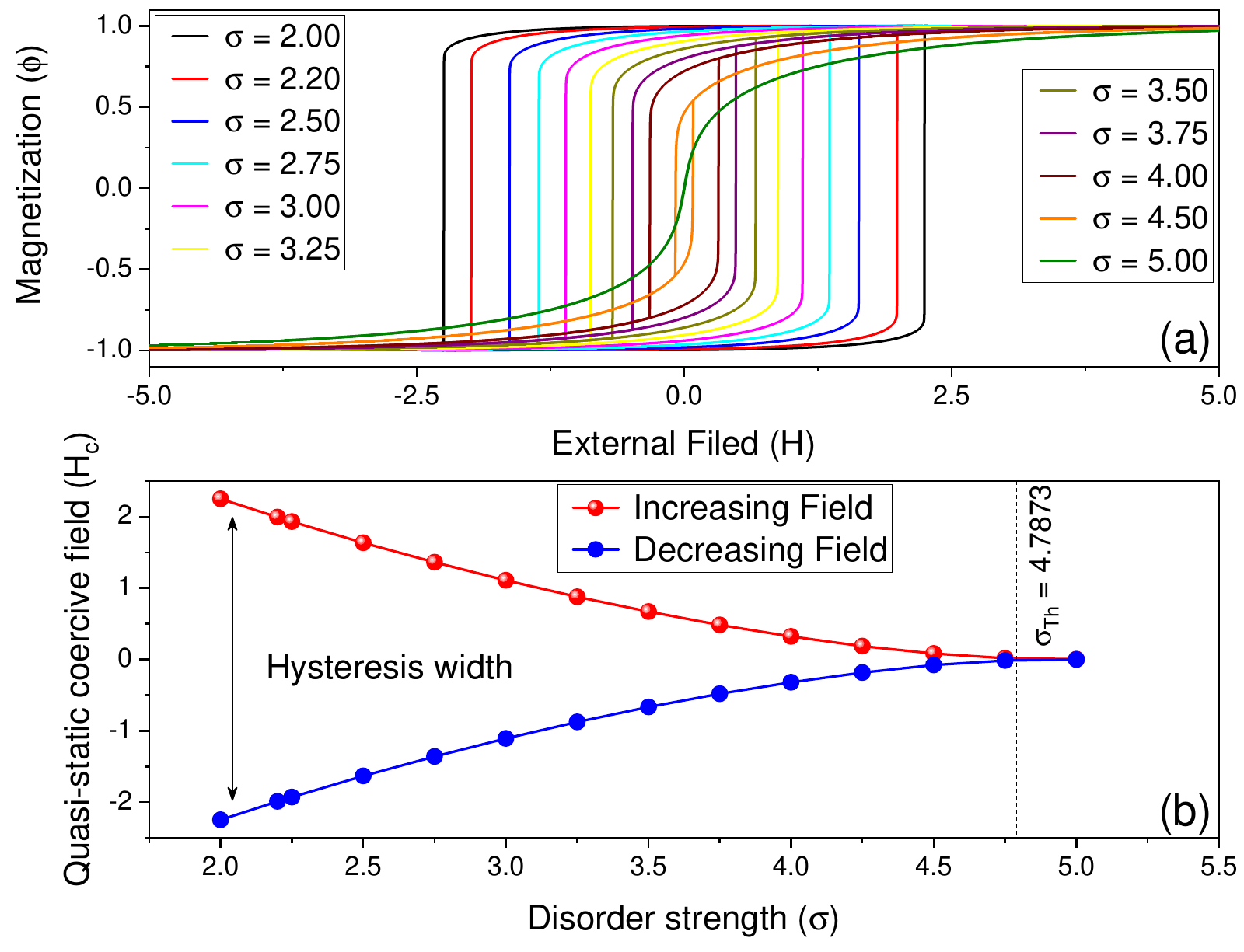}
\caption{(a) Quasi-static magnetization $\phi$ versus external field $H$ for different disorder strengths, obtained from mean-field 3D-ZTRFIM simulations of a system of size $300^3$. The quasi-static process corresponds to an infinitesimal field rate, where the system is allowed to equilibrate at each small field step. The detailed calculation algorithm is described in our previous study~\citep{Bar_prb23}. (b) The hysteresis width decreases with increasing disorder strength and finally vanishes between $\sigma = 4.75$ and $\sigma = 5.00$.}
\label{Hysteresis_width}
\end{figure*}

The non-equilibrium zero-temperature random-field Ising model has long been used to study avalanches, disorder-induced scaling, and hysteresis in athermal systems. The behavior of these phenomena strongly depends on the dimensionality, range of interaction, spin flexibility, and the nature of the dynamics. The quasi-statically driven ZTRFIM has generally been studied using two different types of dynamics. Sethna and co-workers introduced the Metropolis dynamics (or metastable dynamics), which we follow in this work \cite{Sethna_prl93}. In contrast, Robbins and co-workers employed domain-growth dynamics to investigate domain wall motion, where no transition occurs above the critical disorder \cite{Nandi_prl16}.

In the case of metastable dynamics, a power-law distribution of avalanches along with a finite hysteresis width has been observed above the critical point~\cite{Sethna_prl95}. The hysteresis width starts decreasing with the disorder strength and finally vanishes at a higher disorder strength, denoted as $\sigma_{\mathrm{Th}}$. In mean-field calculations, the quasi-static hysteresis of the 3D ZTRFIM vanishes for disorder $\sigma > 4.75$ [Fig.~\ref{Hysteresis_width}]. The exact vanishing point of the hysteresis can be calculated analytically for the mean-field scenario. The value $\sigma_{\mathrm{Th}} = 4.7873$ is consistent with the quasi-static dynamical calculation~\cite{PerezReche_Prb04}. In finite-dimensional calculations (non-mean-field), the hysteresis vanishes at even higher disorder values. In mean field, the transition below $\sigma = 4.7873$ happens at the limit of stability (spinodal), whereas in finite-dimensional calculations the signature of the spinodal is smeared out by local disorder-induced fluctuations~\cite{Bar_prb23}.

Therefore, in the ZTRFIM, saddle-node bifurcation points exist even above the critical disorder $\sigma > \sigma_c$ up to a threshold value of disorder ($\sigma_{\mathrm{Th}}$), where the two branches of the hysteresis curve converge into a single curve. In this region, the catastrophic shift appears in the form of large, system-spanning avalanches and thus it becomes important to predict these transitions before they occur.

\section{Numerical Simulations}

\begin{figure*}
\centering
\includegraphics[width=0.8\textwidth]{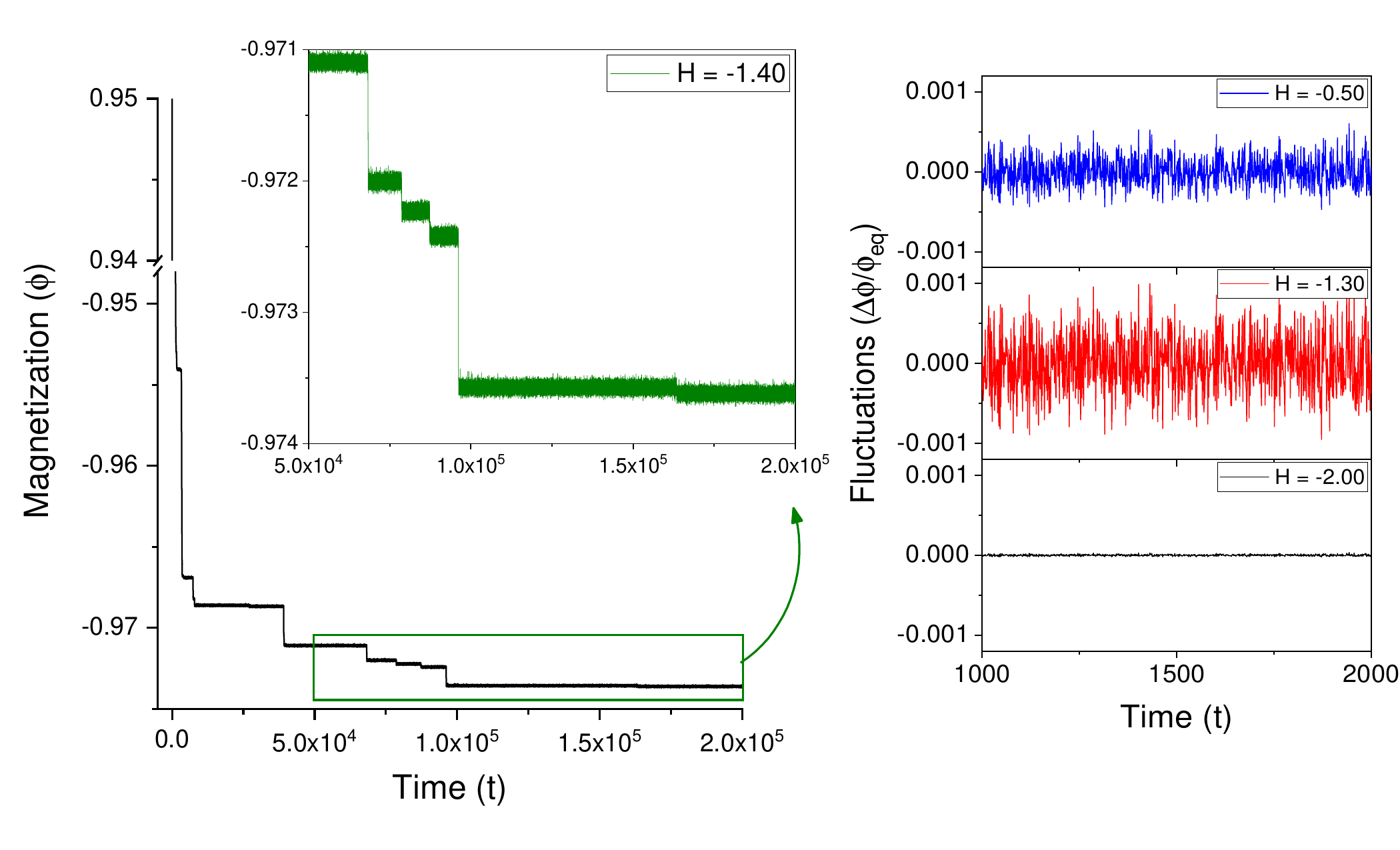}
\caption[0.5\textwidth]{(left) system state (magnetization) as a function of time step. (inset) Zoom-in view of avalanches of different sizes at random times. (right) Time series of the phase fluctuations after reaching the steady state configuration at different external fields.}
\label{fig:Noise}
\end{figure*}

\subsection{Model and method}
This section details the Monte Carlo simulation of the zero temperature random field Ising model (ZTRFIM). The Hamiltonian is given by
\begin{align}
\mathcal{H} = -J \sum_{\langle i,j \rangle} s_i s_j - \sum_{i} \left[ H(t) + h_i \right] s_i,
\label{eqn:RFIM}
\end{align}
The spins at each site can take values, $s_i=\pm 1$, on a three-dimensional (3D) cubic lattice of linear dimension $L$ with periodic boundary conditions \cite{Sethna_prl93, Bar_prb21, Bar_prb23}. The interaction strength $J$ prefers the nearest neighbor spins to align along the same direction. We set $J=1$ for this work. We apply a time-dependent uniform field, $H(t)$, which we change with time. Furthermore, $h_i$ symbolizes a site-specific static magnetic field that varies randomly. Each local random field $h_i$ is drawn from a Gaussian distribution, denoted as $V(h)$, which can be expressed as follows:
\begin{align}
V(h) = \frac{1}{\sqrt{2 \pi \sigma^2}} e^{-h^2/(2\sigma^2)},
\label{eqn:Gaussian}
\end{align}
where $\sigma$ denotes the width of the normal distribution and indicates the disorder strength in this model. We average relevant physical quantities using $100-200$ independent disorder realizations. We perform the calculation at zero temperature, and hence, the spin-flip protocol is determined by the sign change of the local field
\begin{align}
E_i=J\sum_j s_j+ h_i+H, 
\label{eqn:flip}
\end{align}
where the sum over $j$ is the neighbor sites of $i$. The local field completely determines the spin-flip at each time step without thermal/temporal fluctuations.

\subsection{Relaxation time constant}

We quantify the time required for the system to relax to a steady state—via a series of successive avalanches—by analyzing the phase ordering dynamics during small changes in the external field, $\delta H$. Beginning from a fully up-polarized initial state, the external field is gradually reduced in steps of $\delta H$ until the system becomes fully down-polarized. At each field step, the system is allowed to evolve until it reaches a steady state. In numerical simulations, time is typically quantified as the duration of an avalanche, during which all unstable spins from the previous time step are flipped in each iteration \cite{Sethna_prb99}. The relaxation process typically involves a sequence of avalanches. The total time taken to reach the steady state at each field step is defined as the relaxation time constant ($\tau$) for that particular field. The value of $\tau$ is extracted from the temporal evolution of the total magnetization, as explained below.

\begin{enumerate}
\item First, we set up a 3-dimensional integer lattice of system size $L$. At the initial field $H_0 \to +\infty$, the spin at every site is aligned in the up direction ($s_i = 1$). Then, we decrease the external field in steps of $\delta H = 0.0005$ until the system becomes fully down-polarized.

\item At each external field step ($H_f$), we set the time to $t = 0$ and examine every site to determine whether the local field, as defined in Eq.~\ref{eqn:flip}, changes sign.

\item If the local field at one or more sites changes sign, we flip the spin at those sites in the next time step, $t = t + 1$.

\item We then re-check all sites and repeat step 3 until the system reaches a steady state, where no site induces a sign change in the local field.

\item The final time step is considered the relaxation time $\tau$ corresponding to the given quenched field $H_f$.

\item We repeat steps 2–5 for all field values, using the final spin configuration from the previous field $H = H_f - \delta H$ as the initial configuration for the current field $H = H_f$, in order to compute the phase ordering times across the transition region.
\end{enumerate}

The time constant for different fields has been calculated for many disorder configurations, and their averages are presented in Fig. 1(c) of the main text. Here, the time constant represents the total number of successive avalanches and is therefore highly dependent on the external field step size. In our computations, we fix the field step at $\delta H = 0.0005$. Since the time constant is a relative measure, its absolute unit is not meaningful in this context. However, comparing the relative values of the time constant is reasonable for analyzing the system under different disorder strengths.

\subsection{Phase ordering and phase fluctuation}
To comprehend the recovery process following a minor perturbation, we conducted measurements of phase ordering dynamics coupled with a tiny external temporal field fluctuation. First, we start from a fully polarized state and suddenly quench the system to a particular value of the external field. Next, we allow the system to relax when the external field is given by
\begin{align}
    H(t)=H+\eta(t),
\end{align}
where $\eta(t)$ is a  random white (uncorrelated) noise whose amplitude is drawn from a uniform distribution $[-\frac{\eta_0}{2},\frac{\eta_0}{2}]$. We choose $\eta_0= 0.01$, representing the strength of the temporal field fluctuation around the set value $H$. The time $t$ denotes one step where we check the local field of Eq.~(\ref{eqn:flip}) over the whole system for a potential spin flip. Note that in the presence of $\eta(t)$, the system cannot reach a stationary state since the existence of temporal fluctuations can always generate some spin flips.

Figure \ref{fig:Noise} depicts the time-dependent evolution of magnetization starting from a fully polarized up-state, with the field set to $H= -1.4$. The magnetization evolution follows a unidirectional pattern, interspersed with avalanches of varying sizes. Beyond the phase ordering time, these avalanches are triggered by fluctuations in the temporal field, occurring when the total external field $H(t)=H+\eta(t)$ surpasses a specific threshold. It is important to note that the system cannot revert to its previous state once an avalanche occurs, showing the unidirectional nature of athermal systems. As the system reaches a stable configuration, subsequent temporal fluctuations might perturb a few spins without leading to additional avalanches. Instead, the system tends to fluctuate around a specific state.

As a result, fluctuations in the field might introduce noise but cannot trigger an avalanche unless the effective field exceeds a triggering value. Notably, noise, or the fluctuations observed, varies with the specified field strength $H$ [Fig. \ref{fig:Noise}(right)]. At the transition point, there is an increase in noise magnitude, reflecting amplified variance near the tipping point, as depicted in Figure 1(d) in the main text. This amplified noise magnitude is often associated with critical opalescence, an intriguing light-scattering phenomenon observed during continuous phase transitions. However, unlike other long-range interacting systems, the power spectral density of this noise reveals that the fluctuations in the steady state are uncorrelated. Therefore, for a purely athermal condition, the time scale of the system cannot be determined from noise spectroscopic analysis. 

\subsection{Avalanche distribution}

The dynamics of avalanche events in the RFIM for a fixed disorder strongly depend on the driving rate (R) of the external field \cite{Book_Sethna06, Dahmen_prl03, Vives_prl04}. In this section, we discuss the influence of the driving rate on the avalanche distribution. Different driving rates have been incorporated by changing the driving field step size. Here, we follow the \textit{athermal-step-driven} dynamics, where the external field is changed by a certain value and then kept fixed until the system reaches a metastable steady state through successive avalanches \cite{Tadic_prl96, Vives_prl04}.

\begin{figure*}
\centering
\includegraphics[width=0.8\textwidth]{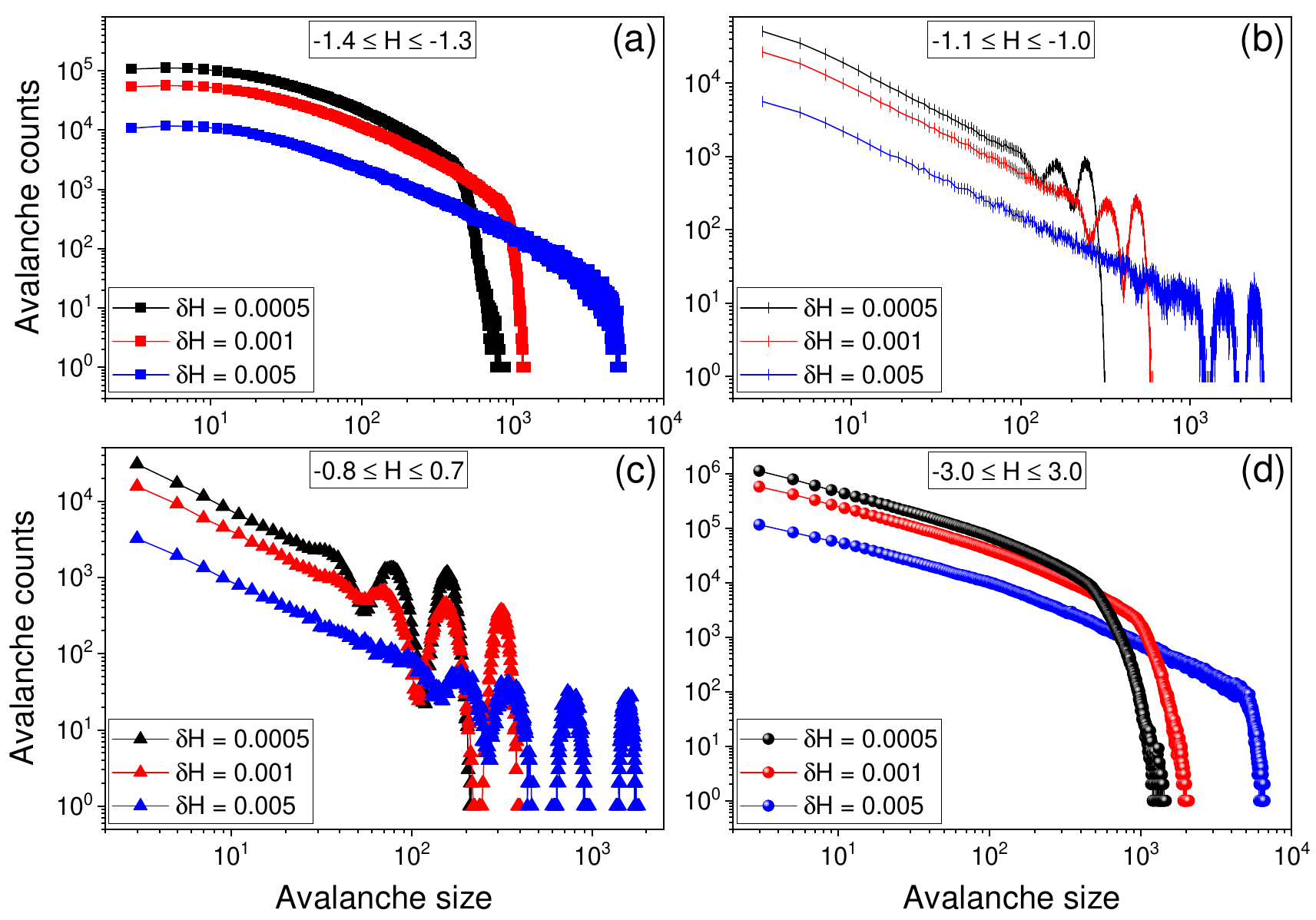}
\caption[0.5\textwidth]{Log-log plots of avalanche size distributions integrated over specified field intervals: (a) near the transition point; (b) and (c) far from the transition point; and (d) over the entire transition region. The simulations were performed with decreasing external fields using three different field steps ($\delta H = 0.005$, $0.001$, $0.0005$), representing three distinct finite driving rates. The simulated results reported here were obtained on a 256$^3$ system size with a disorder strength of $\sigma = 2.3$.}
\label{fig:All_R}
\end{figure*}

Figure \ref{fig:All_R} shows the avalanche distributions for three different driving rates at different field intervals. The distributions shift towards larger-sized avalanches as the driving rate increases. When the field is far from the transition point, the avalanche distribution exhibits several humps towards the larger sizes [Fig. \ref{fig:All_R}(b,c)]. The number of humps decreases with a decreasing driving rate (or step size) and eventually disappears at the adiabatic driving rate (R = 0), where only a single avalanche is allowed to propagate \cite{Book_Sethna06, Dahmen_prl03}. Therefore, the observation of humps in the distribution strongly depends on the driving rate of the external field. Although the avalanche distributions integrated over the full hysteresis branch follow a power-law behavior in the low-avalanche range [Fig. \ref{fig:All_R}(d)], the exponent is heavily dependent on the driving rate. The exponent's value increases as the driving rate (field step) decreases and finally reaches the adiabatic value \cite{Dahmen_prl03, Vives_prl04}.

\subsection{Spatial correlation}
\begin{figure*}
\centering
\includegraphics[width=0.8\textwidth]{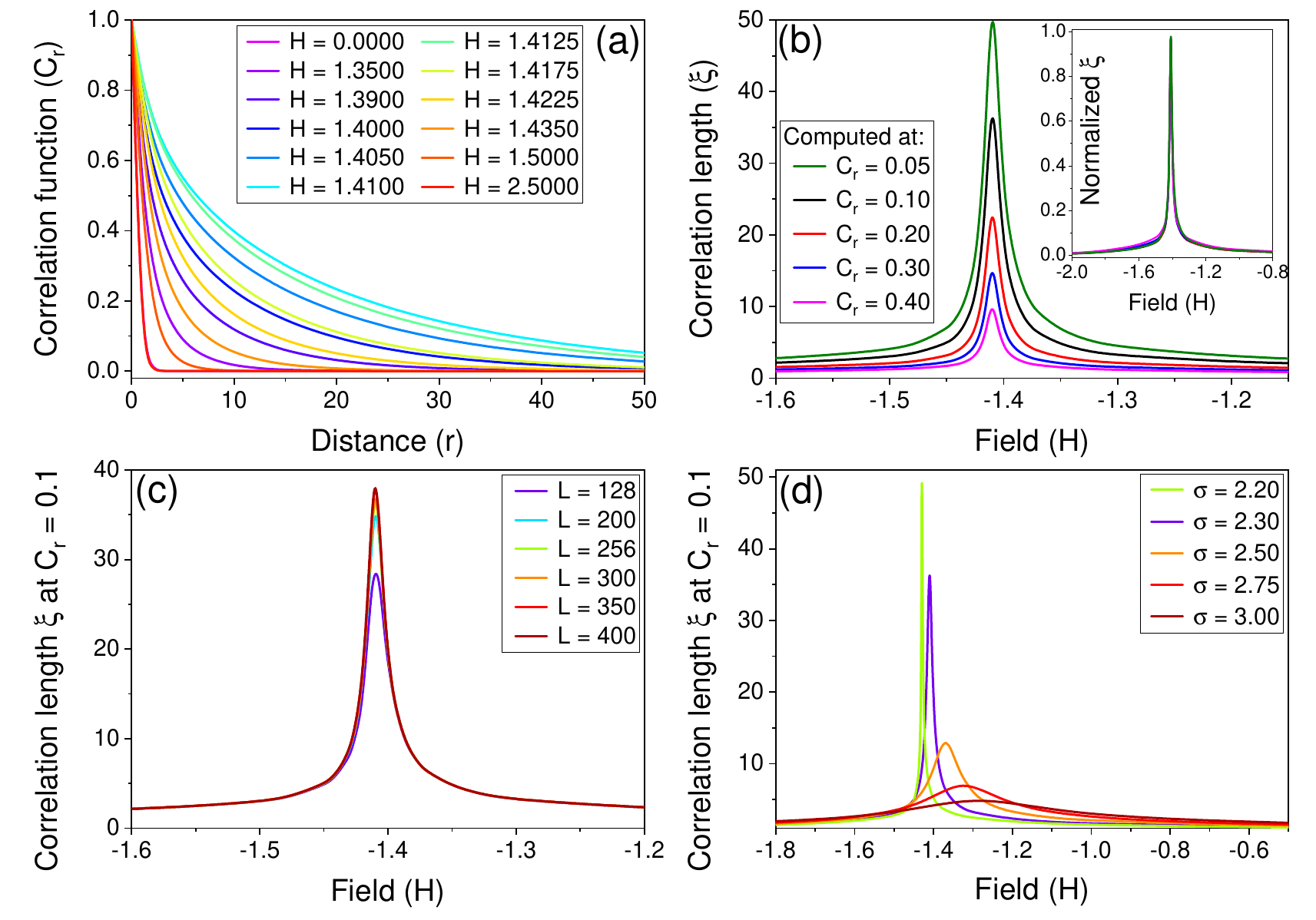}
\caption[0.5\textwidth]{(a) Radial spin-spin correlation functions ($C_r$) at specified external fields. (b) The characteristic correlation length evaluated at different values of $C_r$ expresses the same increasing trend at the transition point. The normalized curves almost coincide altogether (inset). The correlation length versus external field for different system sizes at disorder $\sigma = 2.3$ (c) and for $L=256$ at different disorder strengths (d).}
\label{fig:Correlation}
\end{figure*}

We performed a quasistatic simulation of the ZTRFIM, starting from a fully polarized state, and allowed the system to reach a steady state at each field value before incrementally increasing the field. In the steady-state configuration, we calculate the spin-spin correlation function as a function of their radial distances at each magnetic field. The averaged radial autocorrelations across $100$ disorder configurations are presented in Fig. \ref{fig:Correlation} (a). Notably, the correlation exhibits an increase during the transition phase. Due to the unknown functional form of the correlation function, we compute the correlation length at various segments of the correlation function, as depicted in Fig. \ref{fig:Correlation} (b). The qualitative nature of the correlation length remains consistent irrespective of the arbitrariness of the calculation, showing a coherent alignment in the normalized lengths, as illustrated in the inset of Fig. \ref{fig:Correlation} (b). 

The rising correlation length during the transition is constrained by the total size of smaller systems, exhibiting a finite size effect in the calculations, as illustrated in Figure \ref{fig:Correlation}(c). However, the finite size effect becomes negligible for system sizes $L > 200$. The correlation length at the transition point significantly depends on the disorder strength. As the disorder increases from its critical value, the peak of the correlation length during the transition undergoes both broadening and rapid diminishment [Fig. \ref{fig:Correlation}(d)]. These variations in peak height and width concerning disorder strength are presented in the main text [see Fig. 3(a,b)].

\begin{figure*}
\centering
\includegraphics[width=0.9\textwidth]{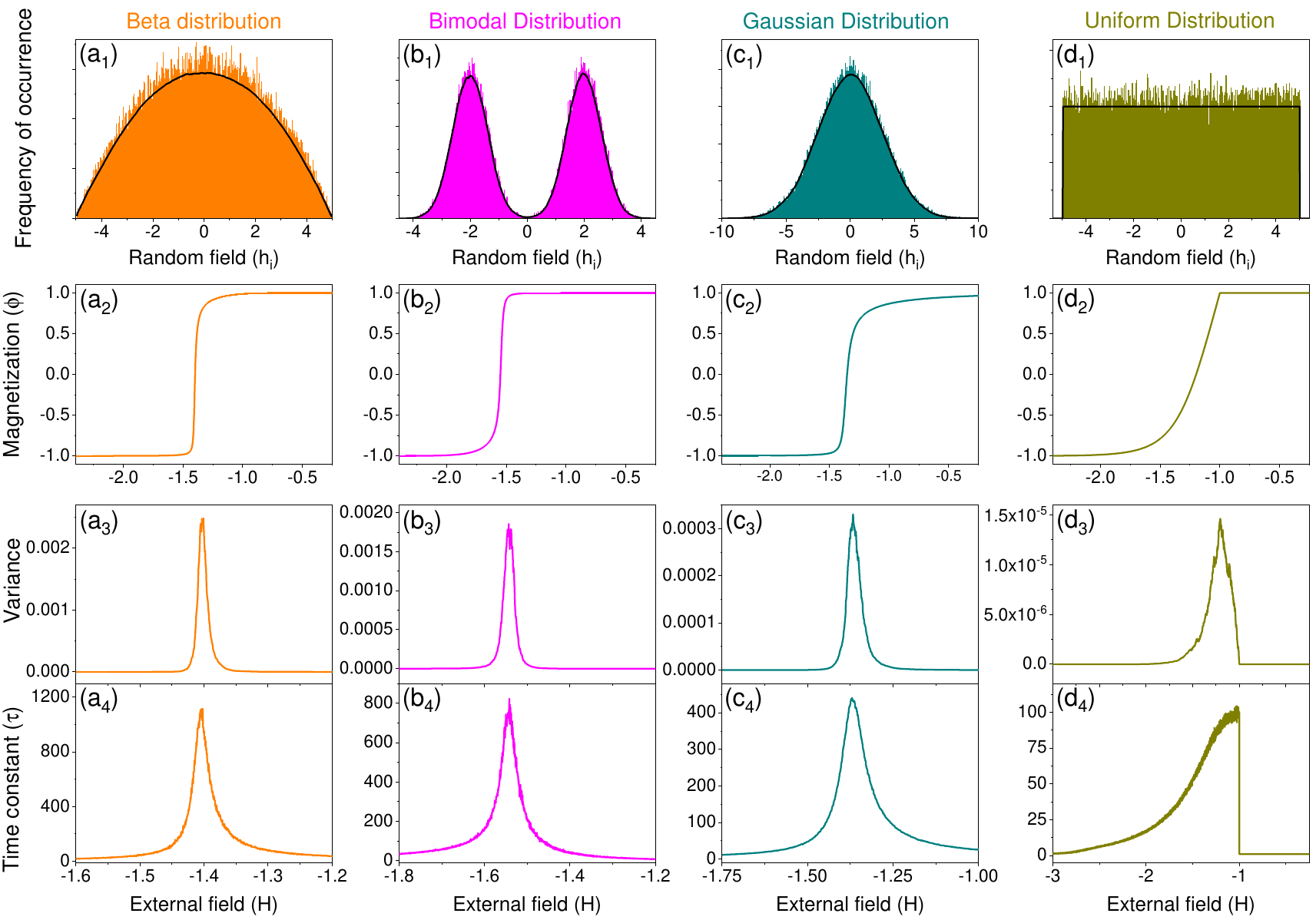}
\caption[0.5\textwidth]{Different disorder distributions (a$_1$,b$_1$,c$_1$,d$_1$) that has been used to cross check the critical transition in the zero temperature random field ising model during slowly decreasing external field ($\Delta H = 0.0005$) and the respective results has been presented column wise. For smooth distributions, the gradual increase in variance and time constant has been observed followed by a cumulative change in order parameter $\phi$ (columns a,b,c). The indicators shows a sudden jumps (d$_3$,d$_4$) accompanied by a sharp twist in the magnetization (d$_2$) in case of uniform distribution. All the data presented here for the simulation done on 256$^3$ system and averaged over 100 disorder configurations.}
\label{fig:All_dist}
\end{figure*}

\begin{figure*}
\centering
\includegraphics[width=0.8\textwidth]{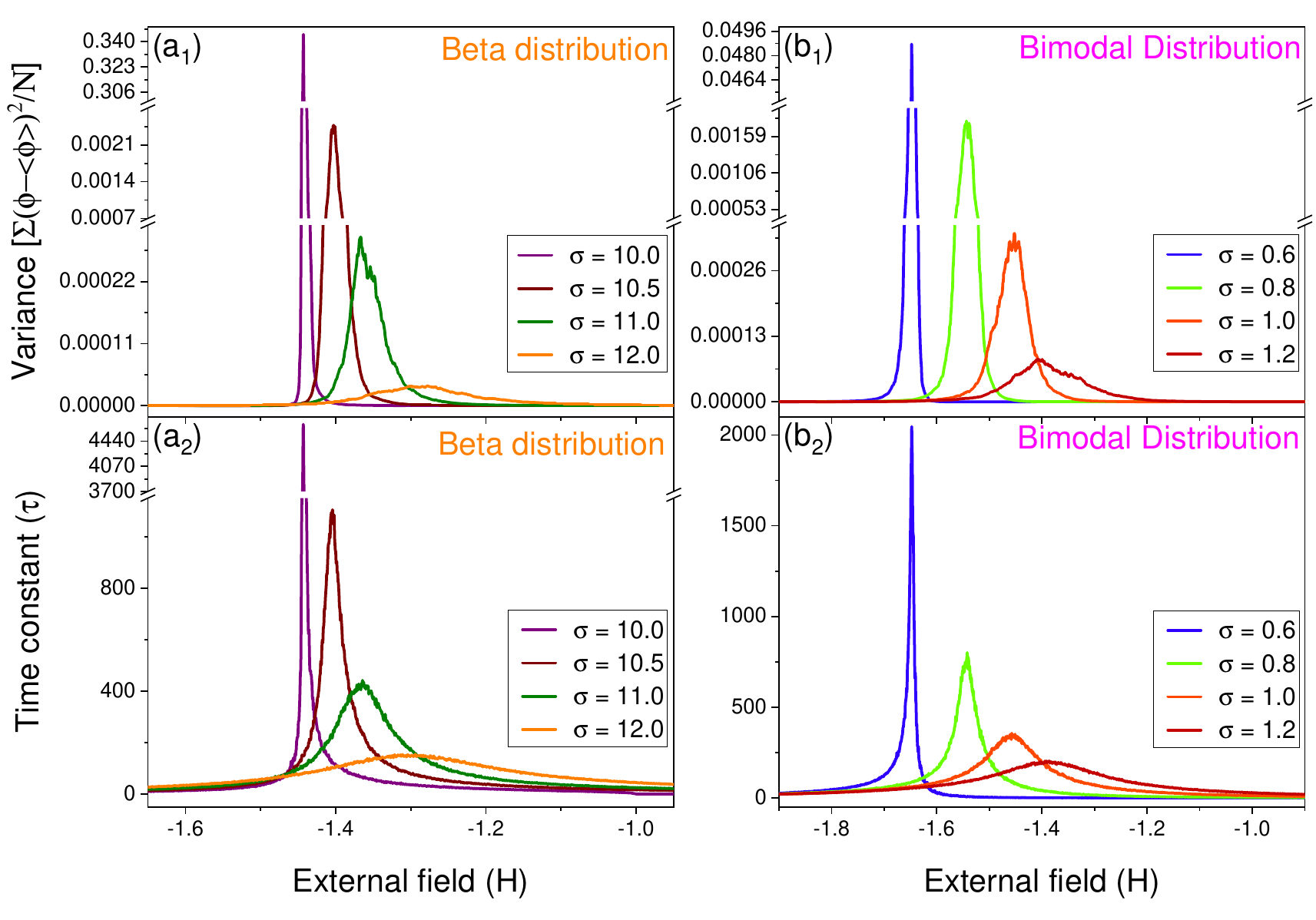}
\caption[0.5\textwidth]{Variance and time constant peak hight decreasing with the increase in disorder strength $\sigma$ for Beta as well as Bimodal disorder configurations. Simultaneously, the peaks width becomes wider for higher disorder.}
\label{fig:Beta_Bimodal}
\end{figure*}

\subsection{Disorder distribution}

In order to provide limitations of the predictions, we perform the random field Ising model simulation with disorders taken from different distributions. The disorder induced critical transition shifted along the disorder line depending upon the distribution of disorder and the way ones define the disorder-strength. In order to compare the numerical results, we first roughly enumerated a disorder value $\sigma_c$ for each and every disorder distribution such that transformation above which becomes successive instead of one step process. Then we carried out all the simulations for disorder strength above the critical values ($\sigma > \sigma_c$).

In Fig.~\ref{fig:All_dist}(a1), the disorder profile is generated from a symmetric Beta distribution,  
\begin{equation}
    f(x) = N x^{a-1}(1-x)^{b-1},
\end{equation}
with parameters $a=b=2$, where $N$ is the normalization constant. To obtain a zero-centered distribution, we shift and rescale the variables so that they take values in the interval $[-1/2,1/2]$. Finally, the rescaled variables are multiplied by a parameter $W$, which controls the overall width of the distribution and therefore the strength of the disorder $\sigma$. The results for the beta distribution shown in Fig.~\ref{fig:All_dist}(a1 to a4) correspond to a disorder strength of $\sigma = 10.5$.

In Fig.~\ref{fig:All_dist}(c1), the disorder profile is generated from a symmetric Gaussian distribution as given in Eq.~\ref{eqn:Gaussian}, with zero mean. Similarly, in Fig~\ref{eqn:doubleGaus}(b1) we employ a bimodal Gaussian distribution, where random numbers are drawn from a mixture of two Gaussians of the form  
\begin{equation}
    V(h) = \frac{1}{\sqrt{2 \pi \sigma^2}} e^{-(h-\mu)^2/(2\sigma^2)}+\frac{1}{\sqrt{2 \pi \sigma^2}} e^{-(h+\mu)^2/(2\sigma^2)},
    \label{eqn:doubleGaus}
\end{equation}
where mean is denoted by $\mu$ and variance $\sigma^2$.  We used $\mu=2$ and $\sigma=0.8$ for Fig.~\ref{fig:All_dist}(b1-b4)

Finally, in Fig.~\ref{fig:All_dist}(d1), the disorder profile is generated from a uniform (box) distribution within the range  
\begin{equation}
    V_i \in \left[-\frac{W}{2}, \frac{W}{2}\right],
\end{equation}
where $W$ defines the disorder strength. Here we set $W=10$.

Figure \ref{fig:All_dist} shows a systematic increase in relaxation time constant and variance of order parameter when the random field $h_i$ (disorder) in the equation \ref{eqn:RFIM} has been fed from Beta, Bimodal or Gaussian distribution. On the contrary, there is a sudden jumps in the time constant and variance once disorder is taken from a linear distribution [fig. \ref{fig:All_dist}(d$_3$,d$_4$)]. The signature of critical slowing down, a continual increase in variance and time constant, for all smooth progression of disorder distributions indicate that the predictions of tipping point through ZTRFIM has a genuine potential. However, strategy would not work for a sharp twisted transformation followed by a abrupt progression of disorder distribution as observed a kink like transition in the case of linear distribution [Fig. \ref{fig:All_dist}(d$_2$)]. A kink at the onset of the transition has long been reported in the Bethe lattice for unimodal random-field distributions \cite{Sabhapandit_JSP00}. In such a disorder distribution, the system does not exhibit any critical signatures before the onset field. The system manifest a sharp increase in time constant and variance [Fig. \ref{fig:All_dist}(d$_3$,d$_4$)] once effective field (external plus random field) hits the tipping point.

To verify the robustness of disorder-aided tipping point predictions, we computed the variance and relaxation time constant for different disorder strength $\sigma$, $\sigma > \sigma_c$, receive from Beta and Bimodal distributions. Figure \ref{fig:Beta_Bimodal} shows that the qualitative nature of disorder-aided early warning signals are in good agreement as observed in the case of Gaussian disorder-distribution [Fig. 3 (main paper)].

\section{Experiments}

\begin{figure*}
\centering
\includegraphics[width=0.8\textwidth]{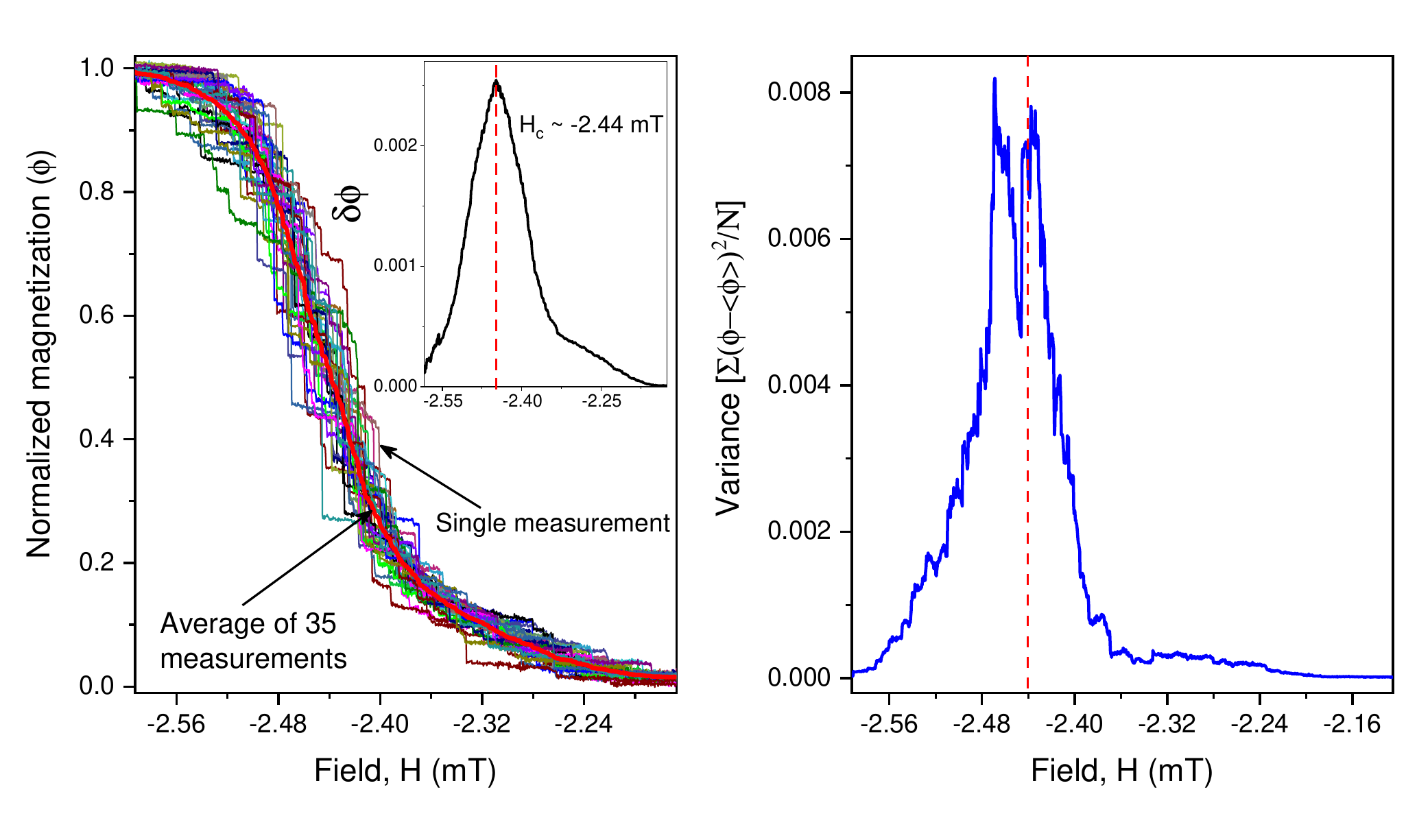}
\caption[0.5\textwidth]{(left) 35 consecutive $\phi-H$ curves and the corresponding average magnetization (red line), $<\phi>$. The data were extracted from the MOKE images, scanned over $5\times 4.3\ mm^2$ Co sample  under decreasing field. (inset) The derivative of average magnetization to extract the coercive field ($H_c = -2.44\ mT$). (right) Variance ($=\sum (\phi - <\phi>)^2/N $) over 35 measurements as a function of decreasing field. Red dotted line pointing the coercive field.}
\label{fig:Variance_Expt}
\end{figure*}

\subsection{Coercive field and variance}

The MOKE images were acquired over a \(5 \times 4.3\ \text{mm}^2\) sample area as a function of decreasing external magnetic field. The fraction of the polarized region (normalized magnetization) is presented in Figure \ref{fig:Variance_Expt} (left) for 35 consecutive measurements. The magnetization exhibits numerous jumps, indicating that the magnetic switching of Co is not a continuous process but occurs through a series of avalanches. The average characteristic behavior of the magnetization (Fig. \ref{fig:Variance_Expt}, solid red line) in a particular sample area can be considered as the total magnetization of a uniform sample in the ergodic limit. Therefore, the derivative of such a transformation can be interpreted as a transition point. 

The discrepancies in magnetization for different consecutive measurements are random, triggered by local dynamic random events. These random events originate from the complex energy barriers that separate several metastable states \cite{Papanikolaou_2018}. In simulations, this process is generally depicted by using different disorder configurations for different runs, and the average results for all disorder configurations fulfill the exact ground state solution \cite{Book_algorithms}. The variation in magnetization at a particular field value for all measurements is treated as the variance of the magnetization. Figure \ref{fig:Variance_Expt} (right) shows the variance of the magnetization over 35 successive measurements. 

The variance may decrease a bit with the number of measurements, but the qualitative nature of the behavior remains unchanged. To obtain accurate numerical values, one can enlarge the configuration space (measurement space); however, the characteristic nature of the variance remains unaffected.

\subsection{Spatial correlation}

In this section, we present the method for extracting the spatial correlation of the MOKE images taken during the decreasing external magnetic field. We have a series of images for different magnetic fields, which change whenever a chunk of spins flip simultaneously (i.e., a new avalanche occurs). Our goal is to determine the 2D autocorrelation of these images.

First, we remove the background by subtracting the first image from all subsequent images. Using the Wiener-Khinchin theorem on the subtracted image, \(X\), the autocorrelation \(C(X)\) is given by:
\begin{equation}
C(X) = F^{-1}[F(X)\tilde{F}(X)]
\end{equation}
where \(F\) and \(F^{-1}\) are the Fourier and inverse Fourier transforms, respectively, and \(\tilde{F}\) represents the complex conjugate of \(F\) \cite{Ryabukho_JO13, VURPILLOT_JM04}.

\begin{figure*}
\centering
\includegraphics[width=1\textwidth]{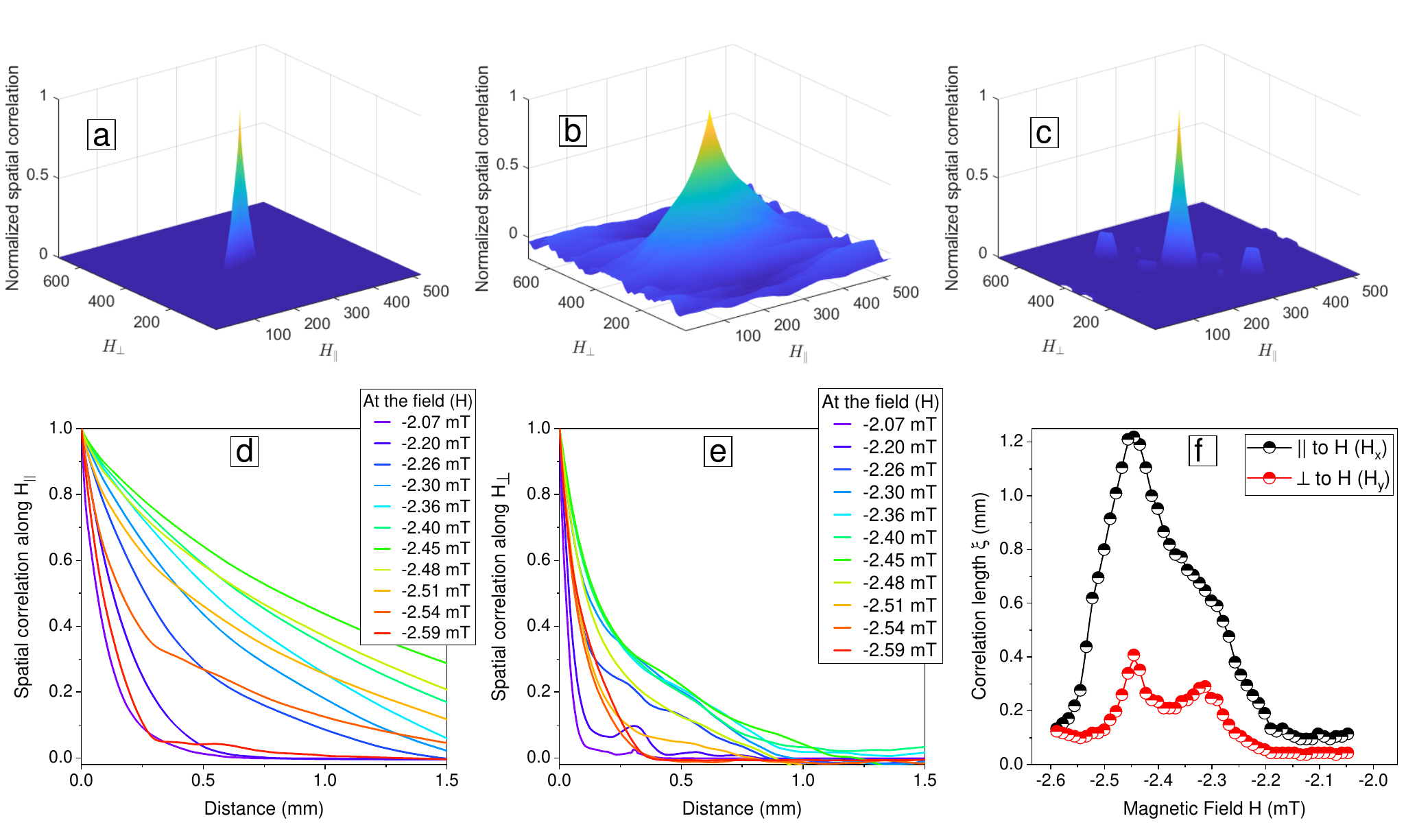}
\caption[0.5\textwidth]{Normalized 2D spatial correlation of $5\times 4.3\ mm^2$ MOKE images (a) before the coercive field, (b) around the coercive field, and (c) after the coercive field. $H_\parallel$ and $H_\perp$ represents the pixel size of the MOKE images parallel and perpendicular to the field directions, respectively. The line of extreme correlation function along (d) $H_\parallel$ and (e) $H_\perp$ as a function of direction for different external fields. (f) The correlation length parallel and perpendicular to the field direction has been extracted at 1/e of the corresponding correlation functions. The data corresponding to the parallel field direction has been presented in the main text.}
\label{fig:Autocorr_Expt}
\end{figure*}

Figure \ref{fig:Autocorr_Expt}(a, b, c) displays the surface plot of the normalized autocorrelation of the MOKE images taken before, during, and after the transition. The images were captured over a \(5 \times 4.3\ \text{mm}^2\) sample area with a sensing area of \(696 \times 520\ \text{pixels}^2\). The increase in correlation around the transition is quite prominent.

The horizontal and vertical distributions of the 2D correlation function for different values of the external magnetic field are presented in Figure \ref{fig:Autocorr_Expt}(d, e). The correlation length is defined as the value at which the correlation function decays to \(1/e\). The obtained correlation length increases as the external field approaches the transformation points.

\bibliography{sample}

\begin{thebibliography}{69}%
\makeatletter
\providecommand \@ifxundefined [1]{%
 \@ifx{#1\undefined}
}%
\providecommand \@ifnum [1]{%
 \ifnum #1\expandafter \@firstoftwo
 \else \expandafter \@secondoftwo
 \fi
}%
\providecommand \@ifx [1]{%
 \ifx #1\expandafter \@firstoftwo
 \else \expandafter \@secondoftwo
 \fi
}%
\providecommand \natexlab [1]{#1}%
\providecommand \enquote  [1]{``#1''}%
\providecommand \bibnamefont  [1]{#1}%
\providecommand \bibfnamefont [1]{#1}%
\providecommand \citenamefont [1]{#1}%
\providecommand \href@noop [0]{\@secondoftwo}%
\providecommand \href [0]{\begingroup \@sanitize@url \@href}%
\providecommand \@href[1]{\@@startlink{#1}\@@href}%
\providecommand \@@href[1]{\endgroup#1\@@endlink}%
\providecommand \@sanitize@url [0]{\catcode `\\12\catcode `\$12\catcode
  `\&12\catcode `\#12\catcode `\^12\catcode `\_12\catcode `\%12\relax}%
\providecommand \@@startlink[1]{}%
\providecommand \@@endlink[0]{}%
\providecommand \url  [0]{\begingroup\@sanitize@url \@url }%
\providecommand \@url [1]{\endgroup\@href {#1}{\urlprefix }}%
\providecommand \urlprefix  [0]{URL }%
\providecommand \Eprint [0]{\href }%
\providecommand \doibase [0]{https://doi.org/}%
\providecommand \selectlanguage [0]{\@gobble}%
\providecommand \bibinfo  [0]{\@secondoftwo}%
\providecommand \bibfield  [0]{\@secondoftwo}%
\providecommand \translation [1]{[#1]}%
\providecommand \BibitemOpen [0]{}%
\providecommand \bibitemStop [0]{}%
\providecommand \bibitemNoStop [0]{.\EOS\space}%
\providecommand \EOS [0]{\spacefactor3000\relax}%
\providecommand \BibitemShut  [1]{\csname bibitem#1\endcsname}%
\let\auto@bib@innerbib\@empty
\bibitem [{\citenamefont {Scheffer}(2020)}]{Book_scheffer2020}%
  \BibitemOpen
  \bibfield  {author} {\bibinfo {author} {\bibfnamefont {M.}~\bibnamefont
  {Scheffer}},\ }\href@noop {} {\emph {\bibinfo {title} {Critical transitions
  in nature and society}}},\ Vol.~\bibinfo {volume} {16}\ (\bibinfo
  {publisher} {Princeton University Press},\ \bibinfo {year}
  {2020})\BibitemShut {NoStop}%
\bibitem [{\citenamefont {Scheffer}\ \emph {et~al.}(2009)\citenamefont
  {Scheffer}, \citenamefont {Bascompte}, \citenamefont {Brock}, \citenamefont
  {Brovkin}, \citenamefont {Carpenter}, \citenamefont {Dakos}, \citenamefont
  {Held}, \citenamefont {van Nes}, \citenamefont {Rietkerk},\ and\
  \citenamefont {Sugihara}}]{Scheffer_Nature09}%
  \BibitemOpen
  \bibfield  {author} {\bibinfo {author} {\bibfnamefont {M.}~\bibnamefont
  {Scheffer}}, \bibinfo {author} {\bibfnamefont {J.}~\bibnamefont {Bascompte}},
  \bibinfo {author} {\bibfnamefont {W.~A.}\ \bibnamefont {Brock}}, \bibinfo
  {author} {\bibfnamefont {V.}~\bibnamefont {Brovkin}}, \bibinfo {author}
  {\bibfnamefont {S.~R.}\ \bibnamefont {Carpenter}}, \bibinfo {author}
  {\bibfnamefont {V.}~\bibnamefont {Dakos}}, \bibinfo {author} {\bibfnamefont
  {H.}~\bibnamefont {Held}}, \bibinfo {author} {\bibfnamefont {E.~H.}\
  \bibnamefont {van Nes}}, \bibinfo {author} {\bibfnamefont {M.}~\bibnamefont
  {Rietkerk}},\ and\ \bibinfo {author} {\bibfnamefont {G.}~\bibnamefont
  {Sugihara}},\ }\bibfield  {title} {\bibinfo {title} {Early-warning signals
  for critical transitions},\ }\href@noop {} {\bibfield  {journal} {\bibinfo
  {journal} {Nature}\ }\textbf {\bibinfo {volume} {461}},\ \bibinfo {pages}
  {53} (\bibinfo {year} {2009})}\BibitemShut {NoStop}%
\bibitem [{\citenamefont {Scheffer}\ \emph {et~al.}(2012)\citenamefont
  {Scheffer}, \citenamefont {Carpenter}, \citenamefont {Lenton}, \citenamefont
  {Bascompte}, \citenamefont {Brock}, \citenamefont {Dakos}, \citenamefont
  {van~de Koppel}, \citenamefont {van~de Leemput}, \citenamefont {Levin},
  \citenamefont {van Nes}, \citenamefont {Pascual},\ and\ \citenamefont
  {Vandermeer}}]{Scheffer_Science12}%
  \BibitemOpen
  \bibfield  {author} {\bibinfo {author} {\bibfnamefont {M.}~\bibnamefont
  {Scheffer}}, \bibinfo {author} {\bibfnamefont {S.~R.}\ \bibnamefont
  {Carpenter}}, \bibinfo {author} {\bibfnamefont {T.~M.}\ \bibnamefont
  {Lenton}}, \bibinfo {author} {\bibfnamefont {J.}~\bibnamefont {Bascompte}},
  \bibinfo {author} {\bibfnamefont {W.}~\bibnamefont {Brock}}, \bibinfo
  {author} {\bibfnamefont {V.}~\bibnamefont {Dakos}}, \bibinfo {author}
  {\bibfnamefont {J.}~\bibnamefont {van~de Koppel}}, \bibinfo {author}
  {\bibfnamefont {I.~A.}\ \bibnamefont {van~de Leemput}}, \bibinfo {author}
  {\bibfnamefont {S.~A.}\ \bibnamefont {Levin}}, \bibinfo {author}
  {\bibfnamefont {E.~H.}\ \bibnamefont {van Nes}}, \bibinfo {author}
  {\bibfnamefont {M.}~\bibnamefont {Pascual}},\ and\ \bibinfo {author}
  {\bibfnamefont {J.}~\bibnamefont {Vandermeer}},\ }\bibfield  {title}
  {\bibinfo {title} {Anticipating critical transitions},\ }\href
  {https://doi.org/10.1126/science.1225244} {\bibfield  {journal} {\bibinfo
  {journal} {Science}\ }\textbf {\bibinfo {volume} {338}},\ \bibinfo {pages}
  {344} (\bibinfo {year} {2012})}\BibitemShut {NoStop}%
\bibitem [{\citenamefont {Lenton}(2011)}]{Lenton_NatCC11}%
  \BibitemOpen
  \bibfield  {author} {\bibinfo {author} {\bibfnamefont {T.~M.}\ \bibnamefont
  {Lenton}},\ }\bibfield  {title} {\bibinfo {title} {Early warning of climate
  tipping points},\ }\href@noop {} {\bibfield  {journal} {\bibinfo  {journal}
  {Nat. Clim. Change}\ }\textbf {\bibinfo {volume} {1}},\ \bibinfo {pages}
  {201–209} (\bibinfo {year} {2011})}\BibitemShut {NoStop}%
\bibitem [{\citenamefont {Lenton}\ \emph {et~al.}(2008)\citenamefont {Lenton},
  \citenamefont {Held}, \citenamefont {Kriegler}, \citenamefont {Hall},
  \citenamefont {Lucht}, \citenamefont {Rahmstorf},\ and\ \citenamefont
  {Schellnhuber}}]{Lenton_PNAS08}%
  \BibitemOpen
  \bibfield  {author} {\bibinfo {author} {\bibfnamefont {T.~M.}\ \bibnamefont
  {Lenton}}, \bibinfo {author} {\bibfnamefont {H.}~\bibnamefont {Held}},
  \bibinfo {author} {\bibfnamefont {E.}~\bibnamefont {Kriegler}}, \bibinfo
  {author} {\bibfnamefont {J.~W.}\ \bibnamefont {Hall}}, \bibinfo {author}
  {\bibfnamefont {W.}~\bibnamefont {Lucht}}, \bibinfo {author} {\bibfnamefont
  {S.}~\bibnamefont {Rahmstorf}},\ and\ \bibinfo {author} {\bibfnamefont
  {H.~J.}\ \bibnamefont {Schellnhuber}},\ }\bibfield  {title} {\bibinfo {title}
  {Tipping elements in the earth's climate system},\ }\href
  {https://doi.org/10.1073/pnas.0705414105} {\bibfield  {journal} {\bibinfo
  {journal} {Proceedings of the National Academy of Sciences}\ }\textbf
  {\bibinfo {volume} {105}},\ \bibinfo {pages} {1786} (\bibinfo {year}
  {2008})}\BibitemShut {NoStop}%
\bibitem [{\citenamefont {Lenton}(2020)}]{Lenton_PTRSB20}%
  \BibitemOpen
  \bibfield  {author} {\bibinfo {author} {\bibfnamefont {T.~M.}\ \bibnamefont
  {Lenton}},\ }\bibfield  {title} {\bibinfo {title} {Tipping positive change},\
  }\href {https://doi.org/10.1098/rstb.2019.0123} {\bibfield  {journal}
  {\bibinfo  {journal} {Philosophical Transactions of the Royal Society B:
  Biological Sciences}\ }\textbf {\bibinfo {volume} {375}},\ \bibinfo {pages}
  {20190123} (\bibinfo {year} {2020})},\ \Eprint
  {https://arxiv.org/abs/https://royalsocietypublishing.org/doi/pdf/10.1098/rstb.2019.0123}
  {https://royalsocietypublishing.org/doi/pdf/10.1098/rstb.2019.0123}
  \BibitemShut {NoStop}%
\bibitem [{\citenamefont {Grimm}(2011)}]{grimm2011predicting}%
  \BibitemOpen
  \bibfield  {author} {\bibinfo {author} {\bibfnamefont {S.}~\bibnamefont
  {Grimm}},\ }\bibfield  {title} {\bibinfo {title} {Predicting social tipping
  points: Current research and the way forward},\ }\href@noop {} {\  (\bibinfo
  {year} {2011})}\BibitemShut {NoStop}%
\bibitem [{\citenamefont {Veraart}\ \emph {et~al.}(2012)\citenamefont
  {Veraart}, \citenamefont {Faassen}, \citenamefont {Dakos}, \citenamefont {van
  Nes}, \citenamefont {Lürling},\ and\ \citenamefont
  {Scheffer}}]{Scheffer1_Nature12}%
  \BibitemOpen
  \bibfield  {author} {\bibinfo {author} {\bibfnamefont {A.~J.}\ \bibnamefont
  {Veraart}}, \bibinfo {author} {\bibfnamefont {E.~J.}\ \bibnamefont
  {Faassen}}, \bibinfo {author} {\bibfnamefont {V.}~\bibnamefont {Dakos}},
  \bibinfo {author} {\bibfnamefont {E.~H.}\ \bibnamefont {van Nes}}, \bibinfo
  {author} {\bibfnamefont {M.}~\bibnamefont {Lürling}},\ and\ \bibinfo
  {author} {\bibfnamefont {M.}~\bibnamefont {Scheffer}},\ }\bibfield  {title}
  {\bibinfo {title} {Recovery rates reflect distance to a tipping point in a
  living system},\ }\href@noop {} {\bibfield  {journal} {\bibinfo  {journal}
  {Nature}\ }\textbf {\bibinfo {volume} {481}},\ \bibinfo {pages} {357}
  (\bibinfo {year} {2012})}\BibitemShut {NoStop}%
\bibitem [{\citenamefont {Dai}\ \emph {et~al.}(2012)\citenamefont {Dai},
  \citenamefont {Vorselen}, \citenamefont {Korolev},\ and\ \citenamefont
  {Gore}}]{Dai_Science12}%
  \BibitemOpen
  \bibfield  {author} {\bibinfo {author} {\bibfnamefont {L.}~\bibnamefont
  {Dai}}, \bibinfo {author} {\bibfnamefont {D.}~\bibnamefont {Vorselen}},
  \bibinfo {author} {\bibfnamefont {K.~S.}\ \bibnamefont {Korolev}},\ and\
  \bibinfo {author} {\bibfnamefont {J.}~\bibnamefont {Gore}},\ }\bibfield
  {title} {\bibinfo {title} {Generic indicators for loss of resilience before a
  tipping point leading to population collapse},\ }\href
  {https://doi.org/10.1126/science.1219805} {\bibfield  {journal} {\bibinfo
  {journal} {Science}\ }\textbf {\bibinfo {volume} {336}},\ \bibinfo {pages}
  {1175} (\bibinfo {year} {2012})}\BibitemShut {NoStop}%
\bibitem [{\citenamefont {Carpenter}\ \emph {et~al.}(2011)\citenamefont
  {Carpenter}, \citenamefont {Cole}, \citenamefont {Pace}, \citenamefont
  {Batt}, \citenamefont {Brock}, \citenamefont {Cline}, \citenamefont {Coloso},
  \citenamefont {Hodgson}, \citenamefont {Kitchell}, \citenamefont {Seekell},
  \citenamefont {Smith},\ and\ \citenamefont {Weidel}}]{Carpenter_Science11}%
  \BibitemOpen
  \bibfield  {author} {\bibinfo {author} {\bibfnamefont {S.~R.}\ \bibnamefont
  {Carpenter}}, \bibinfo {author} {\bibfnamefont {J.~J.}\ \bibnamefont {Cole}},
  \bibinfo {author} {\bibfnamefont {M.~L.}\ \bibnamefont {Pace}}, \bibinfo
  {author} {\bibfnamefont {R.}~\bibnamefont {Batt}}, \bibinfo {author}
  {\bibfnamefont {W.~A.}\ \bibnamefont {Brock}}, \bibinfo {author}
  {\bibfnamefont {T.}~\bibnamefont {Cline}}, \bibinfo {author} {\bibfnamefont
  {J.}~\bibnamefont {Coloso}}, \bibinfo {author} {\bibfnamefont {J.~R.}\
  \bibnamefont {Hodgson}}, \bibinfo {author} {\bibfnamefont {J.~F.}\
  \bibnamefont {Kitchell}}, \bibinfo {author} {\bibfnamefont {D.~A.}\
  \bibnamefont {Seekell}}, \bibinfo {author} {\bibfnamefont {L.}~\bibnamefont
  {Smith}},\ and\ \bibinfo {author} {\bibfnamefont {B.}~\bibnamefont
  {Weidel}},\ }\bibfield  {title} {\bibinfo {title} {Early warnings of regime
  shifts: A whole-ecosystem experiment},\ }\href
  {https://doi.org/10.1126/science.1203672} {\bibfield  {journal} {\bibinfo
  {journal} {Science}\ }\textbf {\bibinfo {volume} {332}},\ \bibinfo {pages}
  {1079} (\bibinfo {year} {2011})}\BibitemShut {NoStop}%
\bibitem [{\citenamefont {Lenton}\ \emph {et~al.}(2012)\citenamefont {Lenton},
  \citenamefont {Livina}, \citenamefont {Dakos}, \citenamefont {Van~Nes},\ and\
  \citenamefont {Scheffer}}]{Lenton_PHRSA2012}%
  \BibitemOpen
  \bibfield  {author} {\bibinfo {author} {\bibfnamefont {T.}~\bibnamefont
  {Lenton}}, \bibinfo {author} {\bibfnamefont {V.}~\bibnamefont {Livina}},
  \bibinfo {author} {\bibfnamefont {V.}~\bibnamefont {Dakos}}, \bibinfo
  {author} {\bibfnamefont {E.}~\bibnamefont {Van~Nes}},\ and\ \bibinfo {author}
  {\bibfnamefont {M.}~\bibnamefont {Scheffer}},\ }\bibfield  {title} {\bibinfo
  {title} {Early warning of climate tipping points from critical slowing down:
  comparing methods to improve robustness},\ }\href@noop {} {\bibfield
  {journal} {\bibinfo  {journal} {Philosophical Transactions of the Royal
  Society A: Mathematical, Physical and Engineering Sciences}\ }\textbf
  {\bibinfo {volume} {370}},\ \bibinfo {pages} {1185} (\bibinfo {year}
  {2012})}\BibitemShut {NoStop}%
\bibitem [{\citenamefont {Zhang}\ \emph {et~al.}(2024)\citenamefont {Zhang},
  \citenamefont {Zhang}, \citenamefont {Liu}, \citenamefont {Zhang},
  \citenamefont {Shao}, \citenamefont {Li}, \citenamefont {Chen}, \citenamefont
  {Liu}, \citenamefont {Ma}, \citenamefont {Han}, \citenamefont {Wang},
  \citenamefont {Adams}, \citenamefont {Shi},\ and\ \citenamefont
  {Ding}}]{Zhang_prl24}%
  \BibitemOpen
  \bibfield  {author} {\bibinfo {author} {\bibfnamefont {J.}~\bibnamefont
  {Zhang}}, \bibinfo {author} {\bibfnamefont {L.-H.}\ \bibnamefont {Zhang}},
  \bibinfo {author} {\bibfnamefont {B.}~\bibnamefont {Liu}}, \bibinfo {author}
  {\bibfnamefont {Z.-Y.}\ \bibnamefont {Zhang}}, \bibinfo {author}
  {\bibfnamefont {S.-Y.}\ \bibnamefont {Shao}}, \bibinfo {author}
  {\bibfnamefont {Q.}~\bibnamefont {Li}}, \bibinfo {author} {\bibfnamefont
  {H.-C.}\ \bibnamefont {Chen}}, \bibinfo {author} {\bibfnamefont {Z.-K.}\
  \bibnamefont {Liu}}, \bibinfo {author} {\bibfnamefont {Y.}~\bibnamefont
  {Ma}}, \bibinfo {author} {\bibfnamefont {T.-Y.}\ \bibnamefont {Han}},
  \bibinfo {author} {\bibfnamefont {Q.-F.}\ \bibnamefont {Wang}}, \bibinfo
  {author} {\bibfnamefont {C.~S.}\ \bibnamefont {Adams}}, \bibinfo {author}
  {\bibfnamefont {B.-S.}\ \bibnamefont {Shi}},\ and\ \bibinfo {author}
  {\bibfnamefont {D.-S.}\ \bibnamefont {Ding}},\ }\bibfield  {title} {\bibinfo
  {title} {Early warning signals of the tipping point in strongly interacting
  rydberg atoms},\ }\href {https://doi.org/10.1103/PhysRevLett.133.243601}
  {\bibfield  {journal} {\bibinfo  {journal} {Phys. Rev. Lett.}\ }\textbf
  {\bibinfo {volume} {133}},\ \bibinfo {pages} {243601} (\bibinfo {year}
  {2024})}\BibitemShut {NoStop}%
\bibitem [{\citenamefont {Gancio}\ and\ \citenamefont
  {Marconi}(2025)}]{Gancio_prl25}%
  \BibitemOpen
  \bibfield  {author} {\bibinfo {author} {\bibfnamefont {M.~C.}\ \bibnamefont
  {Gancio}, \bibfnamefont {Juan}}\ and\ \bibinfo {author} {\bibfnamefont
  {M.}~\bibnamefont {Marconi}},\ }\bibfield  {title} {\bibinfo {title}
  {Identifying and anticipating the threshold bifurcation of a complex laser
  with permutation entropy},\ }\href {https://doi.org/10.1103/8lhf-51kv}
  {\bibfield  {journal} {\bibinfo  {journal} {Phys. Rev. Lett.}\ ,\ } (\bibinfo
  {year} {2025})}\BibitemShut {NoStop}%
\bibitem [{\citenamefont {Morr}\ and\ \citenamefont
  {Boers}(2024)}]{Morr_PRX24}%
  \BibitemOpen
  \bibfield  {author} {\bibinfo {author} {\bibfnamefont {A.}~\bibnamefont
  {Morr}}\ and\ \bibinfo {author} {\bibfnamefont {N.}~\bibnamefont {Boers}},\
  }\bibfield  {title} {\bibinfo {title} {Detection of approaching critical
  transitions in natural systems driven by red noise},\ }\href
  {https://doi.org/10.1103/PhysRevX.14.021037} {\bibfield  {journal} {\bibinfo
  {journal} {Phys. Rev. X}\ }\textbf {\bibinfo {volume} {14}},\ \bibinfo
  {pages} {021037} (\bibinfo {year} {2024})}\BibitemShut {NoStop}%
\bibitem [{\citenamefont {Harris}\ \emph {et~al.}(2024)\citenamefont {Harris},
  \citenamefont {Gollo},\ and\ \citenamefont {Fulcher}}]{Harris_prx24}%
  \BibitemOpen
  \bibfield  {author} {\bibinfo {author} {\bibfnamefont {B.}~\bibnamefont
  {Harris}}, \bibinfo {author} {\bibfnamefont {L.~L.}\ \bibnamefont {Gollo}},\
  and\ \bibinfo {author} {\bibfnamefont {B.~D.}\ \bibnamefont {Fulcher}},\
  }\bibfield  {title} {\bibinfo {title} {Tracking the distance to criticality
  in systems with unknown noise},\ }\href
  {https://doi.org/10.1103/PhysRevX.14.031021} {\bibfield  {journal} {\bibinfo
  {journal} {Phys. Rev. X}\ }\textbf {\bibinfo {volume} {14}},\ \bibinfo
  {pages} {031021} (\bibinfo {year} {2024})}\BibitemShut {NoStop}%
\bibitem [{\citenamefont {Kalimuddin}\ \emph {et~al.}(2024)\citenamefont
  {Kalimuddin}, \citenamefont {Chatterjee}, \citenamefont {Bera}, \citenamefont
  {Afzal}, \citenamefont {Bera}, \citenamefont {Roy}, \citenamefont {Das},
  \citenamefont {Debnath}, \citenamefont {Bansal},\ and\ \citenamefont
  {Mondal}}]{Bansal_prl24}%
  \BibitemOpen
  \bibfield  {author} {\bibinfo {author} {\bibfnamefont {S.}~\bibnamefont
  {Kalimuddin}}, \bibinfo {author} {\bibfnamefont {S.}~\bibnamefont
  {Chatterjee}}, \bibinfo {author} {\bibfnamefont {A.}~\bibnamefont {Bera}},
  \bibinfo {author} {\bibfnamefont {H.}~\bibnamefont {Afzal}}, \bibinfo
  {author} {\bibfnamefont {S.}~\bibnamefont {Bera}}, \bibinfo {author}
  {\bibfnamefont {D.~S.}\ \bibnamefont {Roy}}, \bibinfo {author} {\bibfnamefont
  {S.}~\bibnamefont {Das}}, \bibinfo {author} {\bibfnamefont {T.}~\bibnamefont
  {Debnath}}, \bibinfo {author} {\bibfnamefont {B.}~\bibnamefont {Bansal}},\
  and\ \bibinfo {author} {\bibfnamefont {M.}~\bibnamefont {Mondal}},\
  }\bibfield  {title} {\bibinfo {title} {Exceptionally slow, long-range, and
  non-gaussian critical fluctuations dominate the charge density wave
  transition},\ }\href {https://doi.org/10.1103/PhysRevLett.132.266504}
  {\bibfield  {journal} {\bibinfo  {journal} {Phys. Rev. Lett.}\ }\textbf
  {\bibinfo {volume} {132}},\ \bibinfo {pages} {266504} (\bibinfo {year}
  {2024})}\BibitemShut {NoStop}%
\bibitem [{\citenamefont {Fieguth}(2016)}]{Book_Fieguth2016}%
  \BibitemOpen
  \bibfield  {author} {\bibinfo {author} {\bibfnamefont {P.}~\bibnamefont
  {Fieguth}},\ }\href {https://books.google.es/books?id=sjSgDQAAQBAJ} {\emph
  {\bibinfo {title} {An Introduction to Complex Systems: Society, Ecology, and
  Nonlinear Dynamics}}}\ (\bibinfo  {publisher} {Springer International
  Publishing},\ \bibinfo {year} {2016})\BibitemShut {NoStop}%
\bibitem [{\citenamefont {Garbe}\ \emph {et~al.}(2020)\citenamefont {Garbe},
  \citenamefont {Albrecht}, \citenamefont {Levermann}, \citenamefont {Donges},\
  and\ \citenamefont {Winkelmann}}]{Winkelmann_Nature20}%
  \BibitemOpen
  \bibfield  {author} {\bibinfo {author} {\bibfnamefont {J.}~\bibnamefont
  {Garbe}}, \bibinfo {author} {\bibfnamefont {T.}~\bibnamefont {Albrecht}},
  \bibinfo {author} {\bibfnamefont {A.}~\bibnamefont {Levermann}}, \bibinfo
  {author} {\bibfnamefont {J.~F.}\ \bibnamefont {Donges}},\ and\ \bibinfo
  {author} {\bibfnamefont {R.}~\bibnamefont {Winkelmann}},\ }\bibfield  {title}
  {\bibinfo {title} {The hysteresis of the antarctic ice sheet},\ }\href
  {https://doi.org/10.1038/s41586-020-2727-5} {\bibfield  {journal} {\bibinfo
  {journal} {Nature}\ }\textbf {\bibinfo {volume} {585}},\ \bibinfo {pages}
  {538} (\bibinfo {year} {2020})}\BibitemShut {NoStop}%
\bibitem [{\citenamefont {Akesson}\ \emph {et~al.}(2022)\citenamefont
  {Akesson}, \citenamefont {Morlighem}, \citenamefont {Nilsson}, \citenamefont
  {Stranne},\ and\ \citenamefont {Jakobsson}}]{Akesson_NatCom22}%
  \BibitemOpen
  \bibfield  {author} {\bibinfo {author} {\bibfnamefont {H.}~\bibnamefont
  {Akesson}}, \bibinfo {author} {\bibfnamefont {M.}~\bibnamefont {Morlighem}},
  \bibinfo {author} {\bibfnamefont {J.}~\bibnamefont {Nilsson}}, \bibinfo
  {author} {\bibfnamefont {C.}~\bibnamefont {Stranne}},\ and\ \bibinfo {author}
  {\bibfnamefont {M.}~\bibnamefont {Jakobsson}},\ }\bibfield  {title} {\bibinfo
  {title} {Petermann ice shelf may not recover after a future breakup},\ }\href
  {https://doi.org/10.1038/s41467-022-29529-5} {\bibfield  {journal} {\bibinfo
  {journal} {Nature communications}\ }\textbf {\bibinfo {volume} {13}},\
  \bibinfo {pages} {2519} (\bibinfo {year} {2022})}\BibitemShut {NoStop}%
\bibitem [{\citenamefont {Abe-Ouchi}\ \emph {et~al.}(2013)\citenamefont
  {Abe-Ouchi}, \citenamefont {Saito}, \citenamefont {Kawamura}, \citenamefont
  {Raymo}, \citenamefont {Okuno}, \citenamefont {Takahashi},\ and\
  \citenamefont {Blatter}}]{Abeouchi_Nature13}%
  \BibitemOpen
  \bibfield  {author} {\bibinfo {author} {\bibfnamefont {A.}~\bibnamefont
  {Abe-Ouchi}}, \bibinfo {author} {\bibfnamefont {F.}~\bibnamefont {Saito}},
  \bibinfo {author} {\bibfnamefont {K.}~\bibnamefont {Kawamura}}, \bibinfo
  {author} {\bibfnamefont {M.~E.}\ \bibnamefont {Raymo}}, \bibinfo {author}
  {\bibfnamefont {J.}~\bibnamefont {Okuno}}, \bibinfo {author} {\bibfnamefont
  {K.}~\bibnamefont {Takahashi}},\ and\ \bibinfo {author} {\bibfnamefont
  {H.}~\bibnamefont {Blatter}},\ }\bibfield  {title} {\bibinfo {title}
  {Insolation-driven 100,000-year glacial cycles and hysteresis of ice-sheet
  volume},\ }\href {https://doi.org/10.1038/nature12374} {\bibfield  {journal}
  {\bibinfo  {journal} {nature}\ }\textbf {\bibinfo {volume} {500}},\ \bibinfo
  {pages} {190} (\bibinfo {year} {2013})}\BibitemShut {NoStop}%
\bibitem [{\citenamefont {Sethna}\ \emph {et~al.}(1993)\citenamefont {Sethna},
  \citenamefont {Dahmen}, \citenamefont {Kartha}, \citenamefont {Krumhansl},
  \citenamefont {Roberts},\ and\ \citenamefont {Shore}}]{Sethna_prl93}%
  \BibitemOpen
  \bibfield  {author} {\bibinfo {author} {\bibfnamefont {J.~P.}\ \bibnamefont
  {Sethna}}, \bibinfo {author} {\bibfnamefont {K.}~\bibnamefont {Dahmen}},
  \bibinfo {author} {\bibfnamefont {S.}~\bibnamefont {Kartha}}, \bibinfo
  {author} {\bibfnamefont {J.~A.}\ \bibnamefont {Krumhansl}}, \bibinfo {author}
  {\bibfnamefont {B.~W.}\ \bibnamefont {Roberts}},\ and\ \bibinfo {author}
  {\bibfnamefont {J.~D.}\ \bibnamefont {Shore}},\ }\bibfield  {title} {\bibinfo
  {title} {Hysteresis and hierarchies: Dynamics of disorder-driven first-order
  phase transformations},\ }\href {https://doi.org/10.1103/PhysRevLett.70.3347}
  {\bibfield  {journal} {\bibinfo  {journal} {Phys. Rev. Lett.}\ }\textbf
  {\bibinfo {volume} {70}},\ \bibinfo {pages} {3347} (\bibinfo {year}
  {1993})}\BibitemShut {NoStop}%
\bibitem [{\citenamefont {P\'erez-Reche}\ \emph {et~al.}(2001)\citenamefont
  {P\'erez-Reche}, \citenamefont {Vives}, \citenamefont {Ma\~nosa},\ and\
  \citenamefont {Planes}}]{Planes_prl01}%
  \BibitemOpen
  \bibfield  {author} {\bibinfo {author} {\bibfnamefont {F.~J.}\ \bibnamefont
  {P\'erez-Reche}}, \bibinfo {author} {\bibfnamefont {E.}~\bibnamefont
  {Vives}}, \bibinfo {author} {\bibfnamefont {L.}~\bibnamefont {Ma\~nosa}},\
  and\ \bibinfo {author} {\bibfnamefont {A.}~\bibnamefont {Planes}},\
  }\bibfield  {title} {\bibinfo {title} {Athermal character of structural phase
  transitions},\ }\href {https://doi.org/10.1103/PhysRevLett.87.195701}
  {\bibfield  {journal} {\bibinfo  {journal} {Phys. Rev. Lett.}\ }\textbf
  {\bibinfo {volume} {87}},\ \bibinfo {pages} {195701} (\bibinfo {year}
  {2001})}\BibitemShut {NoStop}%
\bibitem [{\citenamefont {Bar}\ \emph {et~al.}(2021)\citenamefont {Bar},
  \citenamefont {Ghosh},\ and\ \citenamefont {Banerjee}}]{Bar_prb21}%
  \BibitemOpen
  \bibfield  {author} {\bibinfo {author} {\bibfnamefont {T.}~\bibnamefont
  {Bar}}, \bibinfo {author} {\bibfnamefont {A.}~\bibnamefont {Ghosh}},\ and\
  \bibinfo {author} {\bibfnamefont {A.}~\bibnamefont {Banerjee}},\ }\bibfield
  {title} {\bibinfo {title} {Suppression of spinodal instability by disorder in
  an athermal system},\ }\href {https://doi.org/10.1103/PhysRevB.104.144102}
  {\bibfield  {journal} {\bibinfo  {journal} {Phys. Rev. B}\ }\textbf {\bibinfo
  {volume} {104}},\ \bibinfo {pages} {144102} (\bibinfo {year}
  {2021})}\BibitemShut {NoStop}%
\bibitem [{sup()}]{supplementary}%
  \BibitemOpen
  \href@noop {} {\emph {\bibinfo {title} {See the Supplemental Material for the
  details of numerical simulation, experiments and the methods used for data
  analysis \cite{Sethna_prl93, Sethna_prl95, Bar_prl18, Bar_prb21, Bar_prb23,
  Nandi_prl16, Book_algorithms, Book_Nonthermal15, Book_Sethna06,
  Papanikolaou_2018, Ryabukho_JO13, VURPILLOT_JM04, Dahmen_prl03, Tadic_prl96,
  Vives_prl04, PerezReche_Prb04, Sethna_prb99,
  Sabhapandit_JSP00}}}}\BibitemShut {NoStop}%
\bibitem [{\citenamefont {Prettyman}\ \emph {et~al.}(2022)\citenamefont
  {Prettyman}, \citenamefont {Kuna},\ and\ \citenamefont
  {Livina}}]{Prettyman_ERL22}%
  \BibitemOpen
  \bibfield  {author} {\bibinfo {author} {\bibfnamefont {J.}~\bibnamefont
  {Prettyman}}, \bibinfo {author} {\bibfnamefont {T.}~\bibnamefont {Kuna}},\
  and\ \bibinfo {author} {\bibfnamefont {V.}~\bibnamefont {Livina}},\
  }\bibfield  {title} {\bibinfo {title} {Power spectrum scaling as a measure of
  critical slowing down and precursor to tipping points in dynamical systems},\
  }\href {https://doi.org/10.1088/1748-9326/ac526f} {\bibfield  {journal}
  {\bibinfo  {journal} {Environmental Research Letters}\ }\textbf {\bibinfo
  {volume} {17}},\ \bibinfo {pages} {035004} (\bibinfo {year}
  {2022})}\BibitemShut {NoStop}%
\bibitem [{\citenamefont {Abaimov}(2015)}]{Book_Nonthermal15}%
  \BibitemOpen
  \bibfield  {author} {\bibinfo {author} {\bibfnamefont {S.~G.}\ \bibnamefont
  {Abaimov}},\ }\href@noop {} {\emph {\bibinfo {title} {Statistical physics of
  non-thermal phase transitions: from foundations to applications}}}\ (\bibinfo
   {publisher} {Springer},\ \bibinfo {year} {2015})\BibitemShut {NoStop}%
\bibitem [{\citenamefont {Banerjee}\ and\ \citenamefont
  {Bar}(2023)}]{Bar_prb23}%
  \BibitemOpen
  \bibfield  {author} {\bibinfo {author} {\bibfnamefont {A.}~\bibnamefont
  {Banerjee}}\ and\ \bibinfo {author} {\bibfnamefont {T.}~\bibnamefont {Bar}},\
  }\bibfield  {title} {\bibinfo {title} {Finite-dimensional signature of
  spinodal instability in an athermal hysteretic transition},\ }\href
  {https://doi.org/10.1103/PhysRevB.107.024103} {\bibfield  {journal} {\bibinfo
   {journal} {Phys. Rev. B}\ }\textbf {\bibinfo {volume} {107}},\ \bibinfo
  {pages} {024103} (\bibinfo {year} {2023})}\BibitemShut {NoStop}%
\bibitem [{\citenamefont {Bar}\ \emph {et~al.}(2018)\citenamefont {Bar},
  \citenamefont {Choudhary}, \citenamefont {Ashraf}, \citenamefont {Sujith},
  \citenamefont {Puri}, \citenamefont {Raj},\ and\ \citenamefont
  {Bansal}}]{Bar_prl18}%
  \BibitemOpen
  \bibfield  {author} {\bibinfo {author} {\bibfnamefont {T.}~\bibnamefont
  {Bar}}, \bibinfo {author} {\bibfnamefont {S.~K.}\ \bibnamefont {Choudhary}},
  \bibinfo {author} {\bibfnamefont {M.~A.}\ \bibnamefont {Ashraf}}, \bibinfo
  {author} {\bibfnamefont {K.~S.}\ \bibnamefont {Sujith}}, \bibinfo {author}
  {\bibfnamefont {S.}~\bibnamefont {Puri}}, \bibinfo {author} {\bibfnamefont
  {S.}~\bibnamefont {Raj}},\ and\ \bibinfo {author} {\bibfnamefont
  {B.}~\bibnamefont {Bansal}},\ }\bibfield  {title} {\bibinfo {title} {Kinetic
  spinodal instabilities in the mott transition in
  ${\mathrm{v}}_{2}{\mathrm{o}}_{3}$: Evidence from hysteresis scaling and
  dissipative phase ordering},\ }\href
  {https://doi.org/10.1103/PhysRevLett.121.045701} {\bibfield  {journal}
  {\bibinfo  {journal} {Phys. Rev. Lett.}\ }\textbf {\bibinfo {volume} {121}},\
  \bibinfo {pages} {045701} (\bibinfo {year} {2018})}\BibitemShut {NoStop}%
\bibitem [{\citenamefont {Urbach}\ \emph
  {et~al.}(1995{\natexlab{a}})\citenamefont {Urbach}, \citenamefont {Madison},\
  and\ \citenamefont {Markert}}]{Urbach_prl95_athermal}%
  \BibitemOpen
  \bibfield  {author} {\bibinfo {author} {\bibfnamefont {J.~S.}\ \bibnamefont
  {Urbach}}, \bibinfo {author} {\bibfnamefont {R.~C.}\ \bibnamefont
  {Madison}},\ and\ \bibinfo {author} {\bibfnamefont {J.~T.}\ \bibnamefont
  {Markert}},\ }\bibfield  {title} {\bibinfo {title} {Reproducibility of
  magnetic avalanches in an fe-ni-co ferromagnet},\ }\href
  {https://doi.org/10.1103/PhysRevLett.75.4694} {\bibfield  {journal} {\bibinfo
   {journal} {Phys. Rev. Lett.}\ }\textbf {\bibinfo {volume} {75}},\ \bibinfo
  {pages} {4694} (\bibinfo {year} {1995}{\natexlab{a}})}\BibitemShut {NoStop}%
\bibitem [{\citenamefont {Salje}\ \emph {et~al.}(2011)\citenamefont {Salje},
  \citenamefont {Ding}, \citenamefont {Zhao}, \citenamefont {Lookman},\ and\
  \citenamefont {Saxena}}]{Salje_prb11}%
  \BibitemOpen
  \bibfield  {author} {\bibinfo {author} {\bibfnamefont {E.~K.~H.}\
  \bibnamefont {Salje}}, \bibinfo {author} {\bibfnamefont {X.}~\bibnamefont
  {Ding}}, \bibinfo {author} {\bibfnamefont {Z.}~\bibnamefont {Zhao}}, \bibinfo
  {author} {\bibfnamefont {T.}~\bibnamefont {Lookman}},\ and\ \bibinfo {author}
  {\bibfnamefont {A.}~\bibnamefont {Saxena}},\ }\bibfield  {title} {\bibinfo
  {title} {Thermally activated avalanches: Jamming and the progression of
  needle domains},\ }\href {https://doi.org/10.1103/PhysRevB.83.104109}
  {\bibfield  {journal} {\bibinfo  {journal} {Phys. Rev. B}\ }\textbf {\bibinfo
  {volume} {83}},\ \bibinfo {pages} {104109} (\bibinfo {year}
  {2011})}\BibitemShut {NoStop}%
\bibitem [{\citenamefont {Zapperi}\ \emph {et~al.}(1998)\citenamefont
  {Zapperi}, \citenamefont {Cizeau}, \citenamefont {Durin},\ and\ \citenamefont
  {Stanley}}]{Zapperi_prb98}%
  \BibitemOpen
  \bibfield  {author} {\bibinfo {author} {\bibfnamefont {S.}~\bibnamefont
  {Zapperi}}, \bibinfo {author} {\bibfnamefont {P.}~\bibnamefont {Cizeau}},
  \bibinfo {author} {\bibfnamefont {G.}~\bibnamefont {Durin}},\ and\ \bibinfo
  {author} {\bibfnamefont {H.~E.}\ \bibnamefont {Stanley}},\ }\bibfield
  {title} {\bibinfo {title} {Dynamics of a ferromagnetic domain wall:
  Avalanches, depinning transition, and the barkhausen effect},\ }\href
  {https://doi.org/10.1103/PhysRevB.58.6353} {\bibfield  {journal} {\bibinfo
  {journal} {Phys. Rev. B}\ }\textbf {\bibinfo {volume} {58}},\ \bibinfo
  {pages} {6353} (\bibinfo {year} {1998})}\BibitemShut {NoStop}%
\bibitem [{\citenamefont {Shekhawat}\ \emph {et~al.}(2013)\citenamefont
  {Shekhawat}, \citenamefont {Zapperi},\ and\ \citenamefont
  {Sethna}}]{Sethna_prl13}%
  \BibitemOpen
  \bibfield  {author} {\bibinfo {author} {\bibfnamefont {A.}~\bibnamefont
  {Shekhawat}}, \bibinfo {author} {\bibfnamefont {S.}~\bibnamefont {Zapperi}},\
  and\ \bibinfo {author} {\bibfnamefont {J.~P.}\ \bibnamefont {Sethna}},\
  }\bibfield  {title} {\bibinfo {title} {From damage percolation to crack
  nucleation through finite size criticality},\ }\href
  {https://doi.org/10.1103/PhysRevLett.110.185505} {\bibfield  {journal}
  {\bibinfo  {journal} {Phys. Rev. Lett.}\ }\textbf {\bibinfo {volume} {110}},\
  \bibinfo {pages} {185505} (\bibinfo {year} {2013})}\BibitemShut {NoStop}%
\bibitem [{\citenamefont {Sethna}\ \emph {et~al.}(2001)\citenamefont {Sethna},
  \citenamefont {Dahmen},\ and\ \citenamefont {Myers}}]{Sethna_Nature01}%
  \BibitemOpen
  \bibfield  {author} {\bibinfo {author} {\bibfnamefont {J.~P.}\ \bibnamefont
  {Sethna}}, \bibinfo {author} {\bibfnamefont {K.~A.}\ \bibnamefont {Dahmen}},\
  and\ \bibinfo {author} {\bibfnamefont {C.~R.}\ \bibnamefont {Myers}},\
  }\bibfield  {title} {\bibinfo {title} {Crackling noise},\ }\href
  {https://doi.org/10.1038/35065675} {\bibfield  {journal} {\bibinfo  {journal}
  {Nature}\ }\textbf {\bibinfo {volume} {410}},\ \bibinfo {pages} {242}
  (\bibinfo {year} {2001})}\BibitemShut {NoStop}%
\bibitem [{\citenamefont {Perkovi\ifmmode~\acute{c}\else \'{c}\fi{}}\ \emph
  {et~al.}(1995)\citenamefont {Perkovi\ifmmode~\acute{c}\else \'{c}\fi{}},
  \citenamefont {Dahmen},\ and\ \citenamefont {Sethna}}]{Sethna_prl95}%
  \BibitemOpen
  \bibfield  {author} {\bibinfo {author} {\bibfnamefont {O.}~\bibnamefont
  {Perkovi\ifmmode~\acute{c}\else \'{c}\fi{}}}, \bibinfo {author}
  {\bibfnamefont {K.}~\bibnamefont {Dahmen}},\ and\ \bibinfo {author}
  {\bibfnamefont {J.~P.}\ \bibnamefont {Sethna}},\ }\bibfield  {title}
  {\bibinfo {title} {Avalanches, barkhausen noise, and plain old criticality},\
  }\href {https://doi.org/10.1103/PhysRevLett.75.4528} {\bibfield  {journal}
  {\bibinfo  {journal} {Phys. Rev. Lett.}\ }\textbf {\bibinfo {volume} {75}},\
  \bibinfo {pages} {4528} (\bibinfo {year} {1995})}\BibitemShut {NoStop}%
\bibitem [{\citenamefont {White}\ and\ \citenamefont
  {Dahmen}(2003)}]{Dahmen_prl03}%
  \BibitemOpen
  \bibfield  {author} {\bibinfo {author} {\bibfnamefont {R.~A.}\ \bibnamefont
  {White}}\ and\ \bibinfo {author} {\bibfnamefont {K.~A.}\ \bibnamefont
  {Dahmen}},\ }\bibfield  {title} {\bibinfo {title} {Driving rate effects on
  crackling noise},\ }\href {https://doi.org/10.1103/PhysRevLett.91.085702}
  {\bibfield  {journal} {\bibinfo  {journal} {Phys. Rev. Lett.}\ }\textbf
  {\bibinfo {volume} {91}},\ \bibinfo {pages} {085702} (\bibinfo {year}
  {2003})}\BibitemShut {NoStop}%
\bibitem [{\citenamefont {Tadi\ifmmode~\acute{c}\else
  \'{c}\fi{}}(1996)}]{Tadic_prl96}%
  \BibitemOpen
  \bibfield  {author} {\bibinfo {author} {\bibfnamefont {B.}~\bibnamefont
  {Tadi\ifmmode~\acute{c}\else \'{c}\fi{}}},\ }\bibfield  {title} {\bibinfo
  {title} {Nonuniversal scaling behavior of barkhausen noise},\ }\href
  {https://doi.org/10.1103/PhysRevLett.77.3843} {\bibfield  {journal} {\bibinfo
   {journal} {Phys. Rev. Lett.}\ }\textbf {\bibinfo {volume} {77}},\ \bibinfo
  {pages} {3843} (\bibinfo {year} {1996})}\BibitemShut {NoStop}%
\bibitem [{\citenamefont {P\'erez-Reche}\ \emph {et~al.}(2004)\citenamefont
  {P\'erez-Reche}, \citenamefont {Tadi\ifmmode~\acute{c}\else \'{c}\fi{}},
  \citenamefont {Ma\~nosa}, \citenamefont {Planes},\ and\ \citenamefont
  {Vives}}]{Vives_prl04}%
  \BibitemOpen
  \bibfield  {author} {\bibinfo {author} {\bibfnamefont {F.-J.}\ \bibnamefont
  {P\'erez-Reche}}, \bibinfo {author} {\bibfnamefont {B.}~\bibnamefont
  {Tadi\ifmmode~\acute{c}\else \'{c}\fi{}}}, \bibinfo {author} {\bibfnamefont
  {L.}~\bibnamefont {Ma\~nosa}}, \bibinfo {author} {\bibfnamefont
  {A.}~\bibnamefont {Planes}},\ and\ \bibinfo {author} {\bibfnamefont
  {E.}~\bibnamefont {Vives}},\ }\bibfield  {title} {\bibinfo {title} {Driving
  rate effects in avalanche-mediated first-order phase transitions},\ }\href
  {https://doi.org/10.1103/PhysRevLett.93.195701} {\bibfield  {journal}
  {\bibinfo  {journal} {Phys. Rev. Lett.}\ }\textbf {\bibinfo {volume} {93}},\
  \bibinfo {pages} {195701} (\bibinfo {year} {2004})}\BibitemShut {NoStop}%
\bibitem [{\citenamefont {Imry}\ and\ \citenamefont
  {Wortis}(1979)}]{Imry_prb79}%
  \BibitemOpen
  \bibfield  {author} {\bibinfo {author} {\bibfnamefont {Y.}~\bibnamefont
  {Imry}}\ and\ \bibinfo {author} {\bibfnamefont {M.}~\bibnamefont {Wortis}},\
  }\bibfield  {title} {\bibinfo {title} {Influence of quenched impurities on
  first-order phase transitions},\ }\href
  {https://doi.org/10.1103/PhysRevB.19.3580} {\bibfield  {journal} {\bibinfo
  {journal} {Phys. Rev. B}\ }\textbf {\bibinfo {volume} {19}},\ \bibinfo
  {pages} {3580} (\bibinfo {year} {1979})}\BibitemShut {NoStop}%
\bibitem [{\citenamefont {Zohar}\ \emph {et~al.}(2013)\citenamefont {Zohar},
  \citenamefont {Yochelis}, \citenamefont {Dahmen}, \citenamefont {Jung},\ and\
  \citenamefont {Paltiel}}]{Zohar_NatSR13}%
  \BibitemOpen
  \bibfield  {author} {\bibinfo {author} {\bibfnamefont {Y.}~\bibnamefont
  {Zohar}}, \bibinfo {author} {\bibfnamefont {S.}~\bibnamefont {Yochelis}},
  \bibinfo {author} {\bibfnamefont {K.}~\bibnamefont {Dahmen}}, \bibinfo
  {author} {\bibfnamefont {G.}~\bibnamefont {Jung}},\ and\ \bibinfo {author}
  {\bibfnamefont {Y.}~\bibnamefont {Paltiel}},\ }\bibfield  {title} {\bibinfo
  {title} {Controlling avalanche criticality in 2d nano arrays},\ }\href
  {https://doi.org/10.1038/srep01845} {\bibfield  {journal} {\bibinfo
  {journal} {Scientific reports}\ }\textbf {\bibinfo {volume} {3}},\ \bibinfo
  {pages} {1845} (\bibinfo {year} {2013})}\BibitemShut {NoStop}%
\bibitem [{\citenamefont {Laurson}(2024)}]{Laurson_prl24}%
  \BibitemOpen
  \bibfield  {author} {\bibinfo {author} {\bibfnamefont {L.}~\bibnamefont
  {Laurson}},\ }\bibfield  {title} {\bibinfo {title} {Criticality of interface
  depinning and origin of ``bump'' in the avalanche distribution},\ }\href
  {https://doi.org/10.1103/PhysRevLett.133.207102} {\bibfield  {journal}
  {\bibinfo  {journal} {Phys. Rev. Lett.}\ }\textbf {\bibinfo {volume} {133}},\
  \bibinfo {pages} {207102} (\bibinfo {year} {2024})}\BibitemShut {NoStop}%
\bibitem [{\citenamefont {Illa}\ and\ \citenamefont
  {Rosinberg}(2011)}]{Illa_prb11}%
  \BibitemOpen
  \bibfield  {author} {\bibinfo {author} {\bibfnamefont {X.}~\bibnamefont
  {Illa}}\ and\ \bibinfo {author} {\bibfnamefont {M.~L.}\ \bibnamefont
  {Rosinberg}},\ }\bibfield  {title} {\bibinfo {title} {Zero-temperature random
  field ising model on a bethe lattice: Correlation functions along the
  hysteresis loop},\ }\href {https://doi.org/10.1103/PhysRevB.84.064443}
  {\bibfield  {journal} {\bibinfo  {journal} {Phys. Rev. B}\ }\textbf {\bibinfo
  {volume} {84}},\ \bibinfo {pages} {064443} (\bibinfo {year}
  {2011})}\BibitemShut {NoStop}%
\bibitem [{\citenamefont {Handford}\ \emph {et~al.}(2012)\citenamefont
  {Handford}, \citenamefont {Perez-Reche},\ and\ \citenamefont
  {Taraskin}}]{Handford_JSMTE12}%
  \BibitemOpen
  \bibfield  {author} {\bibinfo {author} {\bibfnamefont {T.~P.}\ \bibnamefont
  {Handford}}, \bibinfo {author} {\bibfnamefont {F.-J.}\ \bibnamefont
  {Perez-Reche}},\ and\ \bibinfo {author} {\bibfnamefont {S.~N.}\ \bibnamefont
  {Taraskin}},\ }\bibfield  {title} {\bibinfo {title} {Exact spin–spin
  correlation function for the zero-temperature random-field ising model},\
  }\href {https://doi.org/10.1088/1742-5468/2012/01/P01001} {\bibfield
  {journal} {\bibinfo  {journal} {Journal of Statistical Mechanics: Theory and
  Experiment}\ }\textbf {\bibinfo {volume} {2012}},\ \bibinfo {pages} {P01001}
  (\bibinfo {year} {2012})}\BibitemShut {NoStop}%
\bibitem [{\citenamefont {Perkovi\ifmmode~\acute{c}\else \'{c}\fi{}}\ \emph
  {et~al.}(1999)\citenamefont {Perkovi\ifmmode~\acute{c}\else \'{c}\fi{}},
  \citenamefont {Dahmen},\ and\ \citenamefont {Sethna}}]{Sethna_prb99}%
  \BibitemOpen
  \bibfield  {author} {\bibinfo {author} {\bibfnamefont {O.}~\bibnamefont
  {Perkovi\ifmmode~\acute{c}\else \'{c}\fi{}}}, \bibinfo {author}
  {\bibfnamefont {K.~A.}\ \bibnamefont {Dahmen}},\ and\ \bibinfo {author}
  {\bibfnamefont {J.~P.}\ \bibnamefont {Sethna}},\ }\bibfield  {title}
  {\bibinfo {title} {Disorder-induced critical phenomena in hysteresis:
  Numerical scaling in three and higher dimensions},\ }\href
  {https://doi.org/10.1103/PhysRevB.59.6106} {\bibfield  {journal} {\bibinfo
  {journal} {Phys. Rev. B}\ }\textbf {\bibinfo {volume} {59}},\ \bibinfo
  {pages} {6106} (\bibinfo {year} {1999})}\BibitemShut {NoStop}%
\bibitem [{\citenamefont {Boers}\ and\ \citenamefont
  {Rypdal}(2021)}]{Boers_PNAS21}%
  \BibitemOpen
  \bibfield  {author} {\bibinfo {author} {\bibfnamefont {N.}~\bibnamefont
  {Boers}}\ and\ \bibinfo {author} {\bibfnamefont {M.}~\bibnamefont {Rypdal}},\
  }\bibfield  {title} {\bibinfo {title} {Critical slowing down suggests that
  the western greenland ice sheet is close to a tipping point},\ }\href
  {https://doi.org/10.1073/pnas.2024192118} {\bibfield  {journal} {\bibinfo
  {journal} {Proceedings of the National Academy of Sciences}\ }\textbf
  {\bibinfo {volume} {118}},\ \bibinfo {pages} {e2024192118} (\bibinfo {year}
  {2021})}\BibitemShut {NoStop}%
\bibitem [{\citenamefont {Hirota}\ \emph {et~al.}(2011)\citenamefont {Hirota},
  \citenamefont {Holmgren}, \citenamefont {Nes},\ and\ \citenamefont
  {Scheffer}}]{Scheffer_Science11}%
  \BibitemOpen
  \bibfield  {author} {\bibinfo {author} {\bibfnamefont {M.}~\bibnamefont
  {Hirota}}, \bibinfo {author} {\bibfnamefont {M.}~\bibnamefont {Holmgren}},
  \bibinfo {author} {\bibfnamefont {E.~H.~V.}\ \bibnamefont {Nes}},\ and\
  \bibinfo {author} {\bibfnamefont {M.}~\bibnamefont {Scheffer}},\ }\bibfield
  {title} {\bibinfo {title} {Global resilience of tropical forest and savanna
  to critical transitions},\ }\href {https://doi.org/10.1126/science.1210657}
  {\bibfield  {journal} {\bibinfo  {journal} {Science}\ }\textbf {\bibinfo
  {volume} {334}},\ \bibinfo {pages} {232} (\bibinfo {year}
  {2011})}\BibitemShut {NoStop}%
\bibitem [{\citenamefont {Wang}\ \emph {et~al.}(2012)\citenamefont {Wang},
  \citenamefont {Dearing}, \citenamefont {Langdon}, \citenamefont {Zhang},
  \citenamefont {Yang}, \citenamefont {Dakos},\ and\ \citenamefont
  {Scheffer}}]{Scheffer2_Nature12}%
  \BibitemOpen
  \bibfield  {author} {\bibinfo {author} {\bibfnamefont {R.}~\bibnamefont
  {Wang}}, \bibinfo {author} {\bibfnamefont {J.~A.}\ \bibnamefont {Dearing}},
  \bibinfo {author} {\bibfnamefont {P.~G.}\ \bibnamefont {Langdon}}, \bibinfo
  {author} {\bibfnamefont {E.}~\bibnamefont {Zhang}}, \bibinfo {author}
  {\bibfnamefont {X.}~\bibnamefont {Yang}}, \bibinfo {author} {\bibfnamefont
  {V.}~\bibnamefont {Dakos}},\ and\ \bibinfo {author} {\bibfnamefont
  {M.}~\bibnamefont {Scheffer}},\ }\bibfield  {title} {\bibinfo {title}
  {Flickering gives early warning signals of a critical transition to a
  eutrophic lake state},\ }\href@noop {} {\bibfield  {journal} {\bibinfo
  {journal} {Nature}\ }\textbf {\bibinfo {volume} {492}},\ \bibinfo {pages}
  {419} (\bibinfo {year} {2012})}\BibitemShut {NoStop}%
\bibitem [{\citenamefont {Qiu}\ and\ \citenamefont {Bader}(2000)}]{Qiu_RSI00}%
  \BibitemOpen
  \bibfield  {author} {\bibinfo {author} {\bibfnamefont {Z.~Q.}\ \bibnamefont
  {Qiu}}\ and\ \bibinfo {author} {\bibfnamefont {S.~D.}\ \bibnamefont
  {Bader}},\ }\bibfield  {title} {\bibinfo {title} {{Surface magneto-optic Kerr
  effect}},\ }\href {https://doi.org/10.1063/1.1150496} {\bibfield  {journal}
  {\bibinfo  {journal} {Review of Scientific Instruments}\ }\textbf {\bibinfo
  {volume} {71}},\ \bibinfo {pages} {1243} (\bibinfo {year}
  {2000})}\BibitemShut {NoStop}%
\bibitem [{\citenamefont {Puppin}(2000)}]{Puppin_PRL00}%
  \BibitemOpen
  \bibfield  {author} {\bibinfo {author} {\bibfnamefont {E.}~\bibnamefont
  {Puppin}},\ }\bibfield  {title} {\bibinfo {title} {Statistical properties of
  barkhausen noise in thin fe films},\ }\href
  {https://doi.org/10.1103/PhysRevLett.84.5415} {\bibfield  {journal} {\bibinfo
   {journal} {Phys. Rev. Lett.}\ }\textbf {\bibinfo {volume} {84}},\ \bibinfo
  {pages} {5415} (\bibinfo {year} {2000})}\BibitemShut {NoStop}%
\bibitem [{\citenamefont {Kim}\ \emph {et~al.}(2003)\citenamefont {Kim},
  \citenamefont {Choe},\ and\ \citenamefont {Shin}}]{Kim_PRL03}%
  \BibitemOpen
  \bibfield  {author} {\bibinfo {author} {\bibfnamefont {D.-H.}\ \bibnamefont
  {Kim}}, \bibinfo {author} {\bibfnamefont {S.-B.}\ \bibnamefont {Choe}},\ and\
  \bibinfo {author} {\bibfnamefont {S.-C.}\ \bibnamefont {Shin}},\ }\bibfield
  {title} {\bibinfo {title} {Direct observation of barkhausen avalanche in co
  thin films},\ }\href {https://doi.org/10.1103/PhysRevLett.90.087203}
  {\bibfield  {journal} {\bibinfo  {journal} {Phys. Rev. Lett.}\ }\textbf
  {\bibinfo {volume} {90}},\ \bibinfo {pages} {087203} (\bibinfo {year}
  {2003})}\BibitemShut {NoStop}%
\bibitem [{\citenamefont {Ryu}\ \emph {et~al.}(2007)\citenamefont {Ryu},
  \citenamefont {Akinaga},\ and\ \citenamefont {Shin}}]{Ryu_NatPhys07}%
  \BibitemOpen
  \bibfield  {author} {\bibinfo {author} {\bibfnamefont {K.-S.}\ \bibnamefont
  {Ryu}}, \bibinfo {author} {\bibfnamefont {H.}~\bibnamefont {Akinaga}},\ and\
  \bibinfo {author} {\bibfnamefont {S.-C.}\ \bibnamefont {Shin}},\ }\bibfield
  {title} {\bibinfo {title} {Tunable scaling behaviour observed in barkhausen
  criticality of a ferromagnetic film},\ }\href
  {https://doi.org/10.1038/nphys659} {\bibfield  {journal} {\bibinfo  {journal}
  {Nature Physics}\ }\textbf {\bibinfo {volume} {3}},\ \bibinfo {pages} {547}
  (\bibinfo {year} {2007})}\BibitemShut {NoStop}%
\bibitem [{\citenamefont {Kim}\ \emph {et~al.}(2020)\citenamefont {Kim},
  \citenamefont {Oh}, \citenamefont {Kim}, \citenamefont {Lee}, \citenamefont
  {Baucour}, \citenamefont {Shin}, \citenamefont {Kim}, \citenamefont {Park},\
  and\ \citenamefont {Seo}}]{Kim_NatCom20}%
  \BibitemOpen
  \bibfield  {author} {\bibinfo {author} {\bibfnamefont {D.}~\bibnamefont
  {Kim}}, \bibinfo {author} {\bibfnamefont {Y.-W.}\ \bibnamefont {Oh}},
  \bibinfo {author} {\bibfnamefont {J.~U.}\ \bibnamefont {Kim}}, \bibinfo
  {author} {\bibfnamefont {S.}~\bibnamefont {Lee}}, \bibinfo {author}
  {\bibfnamefont {A.}~\bibnamefont {Baucour}}, \bibinfo {author} {\bibfnamefont
  {J.}~\bibnamefont {Shin}}, \bibinfo {author} {\bibfnamefont {K.-J.}\
  \bibnamefont {Kim}}, \bibinfo {author} {\bibfnamefont {B.-G.}\ \bibnamefont
  {Park}},\ and\ \bibinfo {author} {\bibfnamefont {M.-K.}\ \bibnamefont
  {Seo}},\ }\bibfield  {title} {\bibinfo {title} {Extreme anti-reflection
  enhanced magneto-optic kerr effect microscopy},\ }\href
  {https://doi.org/10.1038/s41467-020-19724-7} {\bibfield  {journal} {\bibinfo
  {journal} {Nature communications}\ }\textbf {\bibinfo {volume} {11}},\
  \bibinfo {pages} {5937} (\bibinfo {year} {2020})}\BibitemShut {NoStop}%
\bibitem [{\citenamefont {Urbach}\ \emph
  {et~al.}(1995{\natexlab{b}})\citenamefont {Urbach}, \citenamefont {Madison},\
  and\ \citenamefont {Markert}}]{Urbach_prl95}%
  \BibitemOpen
  \bibfield  {author} {\bibinfo {author} {\bibfnamefont {J.~S.}\ \bibnamefont
  {Urbach}}, \bibinfo {author} {\bibfnamefont {R.~C.}\ \bibnamefont
  {Madison}},\ and\ \bibinfo {author} {\bibfnamefont {J.~T.}\ \bibnamefont
  {Markert}},\ }\bibfield  {title} {\bibinfo {title} {Interface depinning,
  self-organized criticality, and the barkhausen effect},\ }\href
  {https://doi.org/10.1103/PhysRevLett.75.276} {\bibfield  {journal} {\bibinfo
  {journal} {Phys. Rev. Lett.}\ }\textbf {\bibinfo {volume} {75}},\ \bibinfo
  {pages} {276} (\bibinfo {year} {1995}{\natexlab{b}})}\BibitemShut {NoStop}%
\bibitem [{\citenamefont {Zapperi}\ \emph {et~al.}(1997)\citenamefont
  {Zapperi}, \citenamefont {Ray}, \citenamefont {Stanley},\ and\ \citenamefont
  {Vespignani}}]{Zapperi_prl97}%
  \BibitemOpen
  \bibfield  {author} {\bibinfo {author} {\bibfnamefont {S.}~\bibnamefont
  {Zapperi}}, \bibinfo {author} {\bibfnamefont {P.}~\bibnamefont {Ray}},
  \bibinfo {author} {\bibfnamefont {H.~E.}\ \bibnamefont {Stanley}},\ and\
  \bibinfo {author} {\bibfnamefont {A.}~\bibnamefont {Vespignani}},\ }\bibfield
   {title} {\bibinfo {title} {First-order transition in the breakdown of
  disordered media},\ }\href {https://doi.org/10.1103/PhysRevLett.78.1408}
  {\bibfield  {journal} {\bibinfo  {journal} {Phys. Rev. Lett.}\ }\textbf
  {\bibinfo {volume} {78}},\ \bibinfo {pages} {1408} (\bibinfo {year}
  {1997})}\BibitemShut {NoStop}%
\bibitem [{\citenamefont {Durin}\ and\ \citenamefont
  {Zapperi}(2000)}]{Durin_prl00}%
  \BibitemOpen
  \bibfield  {author} {\bibinfo {author} {\bibfnamefont {G.}~\bibnamefont
  {Durin}}\ and\ \bibinfo {author} {\bibfnamefont {S.}~\bibnamefont
  {Zapperi}},\ }\bibfield  {title} {\bibinfo {title} {Scaling exponents for
  barkhausen avalanches in polycrystalline and amorphous ferromagnets},\ }\href
  {https://doi.org/10.1103/PhysRevLett.84.4705} {\bibfield  {journal} {\bibinfo
   {journal} {Phys. Rev. Lett.}\ }\textbf {\bibinfo {volume} {84}},\ \bibinfo
  {pages} {4705} (\bibinfo {year} {2000})}\BibitemShut {NoStop}%
\bibitem [{\citenamefont {Sethna}\ \emph {et~al.}(2006)\citenamefont {Sethna},
  \citenamefont {Dahmen},\ and\ \citenamefont {Perkovic}}]{Book_Sethna06}%
  \BibitemOpen
  \bibfield  {author} {\bibinfo {author} {\bibfnamefont {J.}~\bibnamefont
  {Sethna}}, \bibinfo {author} {\bibfnamefont {K.}~\bibnamefont {Dahmen}},\
  and\ \bibinfo {author} {\bibfnamefont {O.}~\bibnamefont {Perkovic}},\
  }\bibinfo {title} {Random-field ising models of hysteresis},\ in\ \href
  {https://doi.org/10.1016/B978-012480874-4/50013-0} {\emph {\bibinfo
  {booktitle} {The Science of Hysteresis}}},\ Vol.\ \bibinfo {volume} {2-3}\
  (\bibinfo  {publisher} {Elsevier},\ \bibinfo {year} {2006})\ pp.\ \bibinfo
  {pages} {107--179}\BibitemShut {NoStop}%
\bibitem [{\citenamefont {Im}\ \emph {et~al.}(2009)\citenamefont {Im},
  \citenamefont {Fischer}, \citenamefont {Kim},\ and\ \citenamefont
  {Shin}}]{Im_apl09}%
  \BibitemOpen
  \bibfield  {author} {\bibinfo {author} {\bibfnamefont {M.-Y.}\ \bibnamefont
  {Im}}, \bibinfo {author} {\bibfnamefont {P.}~\bibnamefont {Fischer}},
  \bibinfo {author} {\bibfnamefont {D.-H.}\ \bibnamefont {Kim}},\ and\ \bibinfo
  {author} {\bibfnamefont {S.-C.}\ \bibnamefont {Shin}},\ }\bibfield  {title}
  {\bibinfo {title} {{Direct observation of individual Barkhausen avalanches in
  nucleation-mediated magnetization reversal processes}},\ }\href
  {https://doi.org/10.1063/1.3256188} {\bibfield  {journal} {\bibinfo
  {journal} {Applied Physics Letters}\ }\textbf {\bibinfo {volume} {95}},\
  \bibinfo {pages} {182504} (\bibinfo {year} {2009})}\BibitemShut {NoStop}%
\bibitem [{\citenamefont {Lemerle}\ \emph {et~al.}(1998)\citenamefont
  {Lemerle}, \citenamefont {Ferr\'e}, \citenamefont {Chappert}, \citenamefont
  {Mathet}, \citenamefont {Giamarchi},\ and\ \citenamefont
  {Le~Doussal}}]{Lemerle_prl98}%
  \BibitemOpen
  \bibfield  {author} {\bibinfo {author} {\bibfnamefont {S.}~\bibnamefont
  {Lemerle}}, \bibinfo {author} {\bibfnamefont {J.}~\bibnamefont {Ferr\'e}},
  \bibinfo {author} {\bibfnamefont {C.}~\bibnamefont {Chappert}}, \bibinfo
  {author} {\bibfnamefont {V.}~\bibnamefont {Mathet}}, \bibinfo {author}
  {\bibfnamefont {T.}~\bibnamefont {Giamarchi}},\ and\ \bibinfo {author}
  {\bibfnamefont {P.}~\bibnamefont {Le~Doussal}},\ }\bibfield  {title}
  {\bibinfo {title} {Domain wall creep in an ising ultrathin magnetic film},\
  }\href {https://doi.org/10.1103/PhysRevLett.80.849} {\bibfield  {journal}
  {\bibinfo  {journal} {Phys. Rev. Lett.}\ }\textbf {\bibinfo {volume} {80}},\
  \bibinfo {pages} {849} (\bibinfo {year} {1998})}\BibitemShut {NoStop}%
\bibitem [{\citenamefont {Ashwin}\ \emph {et~al.}(2012)\citenamefont {Ashwin},
  \citenamefont {Wieczorek}, \citenamefont {Vitolo},\ and\ \citenamefont
  {Cox}}]{Ashwin_PHRSA2012}%
  \BibitemOpen
  \bibfield  {author} {\bibinfo {author} {\bibfnamefont {P.}~\bibnamefont
  {Ashwin}}, \bibinfo {author} {\bibfnamefont {S.}~\bibnamefont {Wieczorek}},
  \bibinfo {author} {\bibfnamefont {R.}~\bibnamefont {Vitolo}},\ and\ \bibinfo
  {author} {\bibfnamefont {P.}~\bibnamefont {Cox}},\ }\bibfield  {title}
  {\bibinfo {title} {Tipping points in open systems: bifurcation, noise-induced
  and rate-dependent examples in the climate system},\ }\href
  {https://doi.org/10.1098/rsta.2011.0306} {\bibfield  {journal} {\bibinfo
  {journal} {Philosophical Transactions of the Royal Society A: Mathematical,
  Physical and Engineering Sciences}\ }\textbf {\bibinfo {volume} {370}},\
  \bibinfo {pages} {1166} (\bibinfo {year} {2012})}\BibitemShut {NoStop}%
\bibitem [{\citenamefont {Kundu}\ \emph {et~al.}(2020)\citenamefont {Kundu},
  \citenamefont {Bar}, \citenamefont {Nayak},\ and\ \citenamefont
  {Bansal}}]{Bar_prl20}%
  \BibitemOpen
  \bibfield  {author} {\bibinfo {author} {\bibfnamefont {S.}~\bibnamefont
  {Kundu}}, \bibinfo {author} {\bibfnamefont {T.}~\bibnamefont {Bar}}, \bibinfo
  {author} {\bibfnamefont {R.~K.}\ \bibnamefont {Nayak}},\ and\ \bibinfo
  {author} {\bibfnamefont {B.}~\bibnamefont {Bansal}},\ }\bibfield  {title}
  {\bibinfo {title} {Critical slowing down at the abrupt mott transition: When
  the first-order phase transition becomes zeroth order and looks like second
  order},\ }\href {https://doi.org/10.1103/PhysRevLett.124.095703} {\bibfield
  {journal} {\bibinfo  {journal} {Phys. Rev. Lett.}\ }\textbf {\bibinfo
  {volume} {124}},\ \bibinfo {pages} {095703} (\bibinfo {year}
  {2020})}\BibitemShut {NoStop}%
\bibitem [{\citenamefont {Hagstrom}\ and\ \citenamefont
  {Levin}(2023)}]{Book_Levin23}%
  \BibitemOpen
  \bibfield  {author} {\bibinfo {author} {\bibfnamefont {G.~I.}\ \bibnamefont
  {Hagstrom}}\ and\ \bibinfo {author} {\bibfnamefont {S.~A.}\ \bibnamefont
  {Levin}},\ }\href@noop {} {\emph {\bibinfo {title} {18 Phase Transitions and
  the Theory of Early Warning Indicators for Critical Transitions}}}\ (\bibinfo
   {publisher} {Taylor \& Francis},\ \bibinfo {year} {2023})\ p.\ \bibinfo
  {pages} {358}\BibitemShut {NoStop}%
\bibitem [{\citenamefont {Rundle}\ \emph {et~al.}(2001)\citenamefont {Rundle},
  \citenamefont {Klein}, \citenamefont {Turcotte},\ and\ \citenamefont
  {Malamud}}]{BookChap_Klein01}%
  \BibitemOpen
  \bibfield  {author} {\bibinfo {author} {\bibfnamefont {J.~B.}\ \bibnamefont
  {Rundle}}, \bibinfo {author} {\bibfnamefont {W.}~\bibnamefont {Klein}},
  \bibinfo {author} {\bibfnamefont {D.~L.}\ \bibnamefont {Turcotte}},\ and\
  \bibinfo {author} {\bibfnamefont {B.~D.}\ \bibnamefont {Malamud}},\ }\bibinfo
  {title} {Precursory seismic activation and critical-point phenomena},\ in\
  \href {https://doi.org/10.1007/978-3-0348-7695-7_19} {\emph {\bibinfo
  {booktitle} {Microscopic and Macroscopic Simulation: Towards Predictive
  Modelling of the Earthquake Process}}},\ \bibinfo {editor} {edited by\
  \bibinfo {editor} {\bibfnamefont {P.}~\bibnamefont {Mora}}, \bibinfo {editor}
  {\bibfnamefont {M.}~\bibnamefont {Matsu'ura}}, \bibinfo {editor}
  {\bibfnamefont {R.}~\bibnamefont {Madariaga}},\ and\ \bibinfo {editor}
  {\bibfnamefont {J.-B.}\ \bibnamefont {Minster}}}\ (\bibinfo  {publisher}
  {Birkh{\"a}user Basel},\ \bibinfo {address} {Basel},\ \bibinfo {year}
  {2001})\ pp.\ \bibinfo {pages} {2165--2182}\BibitemShut {NoStop}%
\bibitem [{\citenamefont {Yan}\ \emph {et~al.}(2023)\citenamefont {Yan},
  \citenamefont {Zhang},\ and\ \citenamefont {Wang}}]{Yen_ComPhys23}%
  \BibitemOpen
  \bibfield  {author} {\bibinfo {author} {\bibfnamefont {H.}~\bibnamefont
  {Yan}}, \bibinfo {author} {\bibfnamefont {F.}~\bibnamefont {Zhang}},\ and\
  \bibinfo {author} {\bibfnamefont {J.}~\bibnamefont {Wang}},\ }\bibfield
  {title} {\bibinfo {title} {Thermodynamic and dynamical predictions for
  bifurcations and non-equilibrium phase transitions},\ }\href
  {https://doi.org/10.1038/s42005-023-01210-3} {\bibfield  {journal} {\bibinfo
  {journal} {Communications Physics}\ }\textbf {\bibinfo {volume} {6}},\
  \bibinfo {pages} {110} (\bibinfo {year} {2023})}\BibitemShut {NoStop}%
\bibitem [{\citenamefont {Nandi}\ \emph {et~al.}(2016)\citenamefont {Nandi},
  \citenamefont {Biroli},\ and\ \citenamefont {Tarjus}}]{Nandi_prl16}%
  \BibitemOpen
  \bibfield  {author} {\bibinfo {author} {\bibfnamefont {S.~K.}\ \bibnamefont
  {Nandi}}, \bibinfo {author} {\bibfnamefont {G.}~\bibnamefont {Biroli}},\ and\
  \bibinfo {author} {\bibfnamefont {G.}~\bibnamefont {Tarjus}},\ }\bibfield
  {title} {\bibinfo {title} {Spinodals with disorder: From avalanches in random
  magnets to glassy dynamics},\ }\href
  {https://doi.org/10.1103/PhysRevLett.116.145701} {\bibfield  {journal}
  {\bibinfo  {journal} {Phys. Rev. Lett.}\ }\textbf {\bibinfo {volume} {116}},\
  \bibinfo {pages} {145701} (\bibinfo {year} {2016})}\BibitemShut {NoStop}%
\bibitem [{\citenamefont {P\'erez-Reche}\ and\ \citenamefont
  {Vives}(2004)}]{PerezReche_Prb04}%
  \BibitemOpen
  \bibfield  {author} {\bibinfo {author} {\bibfnamefont {F.~J.}\ \bibnamefont
  {P\'erez-Reche}}\ and\ \bibinfo {author} {\bibfnamefont {E.}~\bibnamefont
  {Vives}},\ }\bibfield  {title} {\bibinfo {title} {Spanning avalanches in the
  three-dimensional gaussian random-field ising model with metastable dynamics:
  Field dependence and geometrical properties},\ }\href
  {https://doi.org/10.1103/PhysRevB.70.214422} {\bibfield  {journal} {\bibinfo
  {journal} {Phys. Rev. B}\ }\textbf {\bibinfo {volume} {70}},\ \bibinfo
  {pages} {214422} (\bibinfo {year} {2004})}\BibitemShut {NoStop}%
\bibitem [{\citenamefont {Sabhapandit}\ \emph {et~al.}(2000)\citenamefont
  {Sabhapandit}, \citenamefont {Shukla},\ and\ \citenamefont
  {Dhar}}]{Sabhapandit_JSP00}%
  \BibitemOpen
  \bibfield  {author} {\bibinfo {author} {\bibfnamefont {S.}~\bibnamefont
  {Sabhapandit}}, \bibinfo {author} {\bibfnamefont {P.}~\bibnamefont
  {Shukla}},\ and\ \bibinfo {author} {\bibfnamefont {D.}~\bibnamefont {Dhar}},\
  }\bibfield  {title} {\bibinfo {title} {Distribution of avalanche sizes in the
  hysteretic response of the random-field ising model on a bethe lattice at
  zero temperature},\ }\href {https://doi.org/10.1023/A:1018622805347}
  {\bibfield  {journal} {\bibinfo  {journal} {Journal of Statistical Physics}\
  }\textbf {\bibinfo {volume} {98}},\ \bibinfo {pages} {103} (\bibinfo {year}
  {2000})}\BibitemShut {NoStop}%
\bibitem [{\citenamefont {Papanikolaou}\ \emph {et~al.}(2017)\citenamefont
  {Papanikolaou}, \citenamefont {Cui},\ and\ \citenamefont
  {Ghoniem}}]{Papanikolaou_2018}%
  \BibitemOpen
  \bibfield  {author} {\bibinfo {author} {\bibfnamefont {S.}~\bibnamefont
  {Papanikolaou}}, \bibinfo {author} {\bibfnamefont {Y.}~\bibnamefont {Cui}},\
  and\ \bibinfo {author} {\bibfnamefont {N.}~\bibnamefont {Ghoniem}},\
  }\bibfield  {title} {\bibinfo {title} {Avalanches and plastic flow in crystal
  plasticity: an overview},\ }\href {https://doi.org/10.1088/1361-651X/aa97ad}
  {\bibfield  {journal} {\bibinfo  {journal} {Modelling and Simulation in
  Materials Science and Engineering}\ }\textbf {\bibinfo {volume} {26}},\
  \bibinfo {pages} {013001} (\bibinfo {year} {2017})}\BibitemShut {NoStop}%
\bibitem [{\citenamefont {Hartmann}\ and\ \citenamefont
  {Rieger}(2002)}]{Book_algorithms}%
  \BibitemOpen
  \bibfield  {author} {\bibinfo {author} {\bibfnamefont {A.~K.}\ \bibnamefont
  {Hartmann}}\ and\ \bibinfo {author} {\bibfnamefont {H.}~\bibnamefont
  {Rieger}},\ }\href@noop {} {\emph {\bibinfo {title} {Optimization algorithms
  in physics}}}\ (\bibinfo  {publisher} {WILEY-VCH Verlag, Berlin},\ \bibinfo
  {year} {2002})\BibitemShut {NoStop}%
\bibitem [{\citenamefont {Ryabukho}\ \emph {et~al.}(2013)\citenamefont
  {Ryabukho}, \citenamefont {Lyakin}, \citenamefont {Grebenyuk},\ and\
  \citenamefont {Klykov}}]{Ryabukho_JO13}%
  \BibitemOpen
  \bibfield  {author} {\bibinfo {author} {\bibfnamefont {V.~P.}\ \bibnamefont
  {Ryabukho}}, \bibinfo {author} {\bibfnamefont {D.~V.}\ \bibnamefont
  {Lyakin}}, \bibinfo {author} {\bibfnamefont {A.~A.}\ \bibnamefont
  {Grebenyuk}},\ and\ \bibinfo {author} {\bibfnamefont {S.~S.}\ \bibnamefont
  {Klykov}},\ }\bibfield  {title} {\bibinfo {title} {Wiener–khintchin theorem
  for spatial coherence of optical wave field},\ }\href
  {https://doi.org/10.1088/2040-8978/15/2/025405} {\bibfield  {journal}
  {\bibinfo  {journal} {Journal of Optics}\ }\textbf {\bibinfo {volume} {15}},\
  \bibinfo {pages} {025405} (\bibinfo {year} {2013})}\BibitemShut {NoStop}%
\bibitem [{\citenamefont {Vurpillot}\ \emph {et~al.}(2004)\citenamefont
  {Vurpillot}, \citenamefont {de~Geuser}, \citenamefont {da~Costa},\ and\
  \citenamefont {Blavette}}]{VURPILLOT_JM04}%
  \BibitemOpen
  \bibfield  {author} {\bibinfo {author} {\bibfnamefont {F.}~\bibnamefont
  {Vurpillot}}, \bibinfo {author} {\bibfnamefont {F.}~\bibnamefont
  {de~Geuser}}, \bibinfo {author} {\bibfnamefont {G.}~\bibnamefont
  {da~Costa}},\ and\ \bibinfo {author} {\bibfnamefont {D.}~\bibnamefont
  {Blavette}},\ }\bibfield  {title} {\bibinfo {title} {Application of fourier
  transform and autocorrelation to cluster identification in the
  three-dimensional atom probe},\ }\href
  {https://doi.org/https://doi.org/10.1111/j.0022-2720.2004.01413.x} {\bibfield
   {journal} {\bibinfo  {journal} {Journal of Microscopy}\ }\textbf {\bibinfo
  {volume} {216}},\ \bibinfo {pages} {234} (\bibinfo {year}
  {2004})}\BibitemShut {NoStop}%
\end{thebibliography}%
\end{document}